\documentclass[]{aa}    
\usepackage{graphicx}

\begin{document}

\title{NGC~4654: polarized radio continuum emission as a diagnostic tool for a galaxy--cluster interaction}
\subtitle{Models versus observations}
\author{M. Soida,  \inst{1} K. Otmianowska-Mazur, \inst{1}  
K.  Chy\.zy \inst{1}
\and
B. Vollmer \inst{2}
}
\institute{Astronomical Observatory,
	   Jagiellonian University, Orla 171,
   30-244 Krak\'ow, Poland 
\and
CDS, Observatoire Astronomique de Strasbourg, UMR 7550, 11, rue de
l'universit\'e, 67000 Strasbourg, France} 

\date{Received }

\abstract
{A recent comparison between deep VLA H{\sc i} observations and dynamical
models of the Virgo cluster spiral galaxy NGC~4654 has shown that only a model
involving a combination of a tidal interaction and ram pressure can
reproduce the data.}
{Deep radio polarization studies, together with detailed MHD modeling, 
can independently verify those conclusions, that are based on H{\sc i}
observations and dynamical models.}
{We performed deep polarized radio-continuum observations of the Virgo
cluster spiral galaxy
NGC~4654 with the Effelsberg 100m telescope at 8.35~GHz and the VLA at
4.85~GHz. Detailed 3D MHD simulations were made to
determine the large-scale magnetic field and the emission distribution of the
polarized radio continuum in the model, during the galaxy evolution within the
cluster environment.}
{This direct comparison between the observed and simulated
polarized radio continuum emission corroborates the earlier results,
that the galaxy had a recent rapid close encounter with NGC~4639 and is
undergoing weak ram pressure by the intracluster medium.}
{This combination of deep radio polarization studies and detailed MHD modeling 
thus gives us unique insight into the interactions of a galaxy with its
cluster environment. It represents a diagnostic tool that is complementary
to deep H{\sc i} observations.}


\authorrunning{Soida et al.}
\titlerunning{Polarized radio continuum emission of NGC~4654}

\maketitle

\section{Introduction \label{sec:introduction}}

The bright Virgo cluster spiral galaxy NGC~4654 is located at the periphery of the
cluster at a projected distance of $3.4^{\circ}=1$~Mpc\footnote{We use
a distance of 17~Mpc to the Virgo cluster} from the cluster center (M87).
Its two-armed spiral structure is asymmetric with a shorter spiral arm to the
northwest and a larger wider spiral arm to the southeast. Its H{\sc i}
distribution shows a sharp edge to the northwest and a large, very low column
density tail to the southeast (Phookun \& Mundy \cite{pm95}).
While the northwestern part of the H{\sc i} velocity field corresponds to a rising
rotation curve, rotation is almost absent in the very low column-density tail.
The H$\alpha$ image (Koopmann et al. \cite{koo01}) shows enhanced star formation at
the northwestern edge of the galactic disk.

Another spiral galaxy, NGC~4639, is located only $18'=90$~kpc in projection from NGC~4654
and might be considered to be a companion galaxy of NGC~4654.
NGC~4639 has a ring-like H{\sc i} distribution. Since it is located at the edge
of the field observed by Phookun \& Mundy (\cite{pm95}) and
Cayatte et al.~(\cite{cay94}), 
it is not clear if its gas distribution and kinematics are perturbed. 
While Warmels (\cite{war88}) observed extended gas in the east of NGC~4639, 
Cayatte et al. (\cite{cay94}) found atomic gas to the southwest. 

Two kinds of interactions can in principle be responsible for the observed
asymmetries and perturbations of NGC~4654: (i) a gravitational interaction 
with a massive galaxy or (ii) ram pressure due to the galaxy's rapid motion
within the hot and tenuous intracluster medium.
In order to investigate the recent evolutionary history of NGC~4654, 
Vollmer (\cite{vol03}) made numerical simulations including a gravitational interaction, 
ram pressure, and a combination of both, concluding that 
\begin{itemize}
\item
a past ram-pressure stripping event where the galaxy has already passed 
the cluster center 800~Myr ago and is now emerging from the cluster core gives
rise to a low surface-density tail in the southeast of the galaxy center.
This tail shows clear signs of rotation;
\item
a gravitational interaction between NGC~4654 and NGC~4639 cannot produce a
low surface density tail;
\item
only a gravitational together with a relatively small amount of ram pressure
leads to the observed low surface-density tail where rotation is almost absent.
\end{itemize}

The results of Vollmer (\cite{vol03}) suggest that NGC~4654 had a tidal interaction 
with another massive galaxy in the recent past ($\sim$500~Myr ago).
A model where the position and radial velocity of the perturbing massive galaxy 
are reasonably close to those of NGC~4639 can explain the perturbed stellar
distribution and the gas distribution of the inner disk.
Therefore, Vollmer (\cite{vol03}) identified NGC~4639 as the perturbing galaxy.

A caveat of the gravitational interaction scenario is that NIR
Tully-Fisher measurements yield a line-of-sight distance between NGC~4654 and NGC~4639
of more than $5$~Mpc (Gavazzi et al. \cite{gav99}).
While NGC~4654 is located $1-2$~Mpc in front of the cluster core,
NGC~4639 is located far behind ($\sim 5\pm1$~Mpc). 
Observations of Cepheids give independent distance estimates for a
number of Virgo cluster galaxies (Gibson et al. \cite{gib00}, Freedman et al. \cite{fre01}).
Five galaxies are in both, the Gavazzi et al.~(\cite{gav99}) and
the Freedman et al.~(\cite{fre01}) samples.
For 4 galaxies (NGC~4321, NGC~4535, NGC~4536, and NGC~4536),
the difference between the distance moduli obtained from the two
methods are small ($|\Delta \mu_0| < 0.2$). However, this difference is much larger for
NGC~4639 ($|\Delta \mu_0| =0.4$). The Cepheid distance estimates place NGC~4639 closer,
but still a few Mpc behind the center of the Virgo cluster (M87).
Thus these measurements corroborate the large NIR Tully Fisher distance of NGC~4639.

Since the preferred scenario of Vollmer et al. (\cite{vol03}) includes 
ram pressure of $p_{\rm ram}=200$~cm$^{-3}$(km\,s$^{-1}$)$^{-1}$, a
galaxy velocity of $\sim 1000$~km\,s$^{-1}$ implies an intracluster medium (ICM)
density of $n_{\rm ICM} \sim 10^{-4}$~cm$^{-3}$. At the distance of NGC~4639,
the ICM density according to Schindler et al. (\cite{sch99})
is lower than $10^{-5}$~cm$^{-3}$.
This suggests that
either (i) NGC~4654 had a gravitational interaction
with another massive object, (ii) the Tully Fisher method  
underestimates the line-of-sight position of NGC~4654, (i.e. NGC~4654
is too bright with respect to its rotation velocity),
or (iii) the ICM distribution is elongated along the line-of-sight,
similar to the galaxy distribution (Gavazzi et al. \cite{gav99}).
In the case of scenario (i) the perturber has to be a massive dark galaxy
(see, e.g., Minchin et al. \cite{min05}) since there is no other visible companion
than NGC~4639.

In this article we test the viability of the preferred scenario of Vollmer (\cite{vol03}).
We used polarized radio-continuum observations of NGC~4654,
together with an MHD model, to put further constraints on the models.
The radio continuum emission is due to relativistic electrons that
gyrate around the galactic magnetic field (synchrotron emission).
If the polarized radio-continuum emission (called PRCE hereafter)
is free of Faraday rotation 
(typically at frequencies above $\sim 4$~GHz), it traces the ordered
large-scale magnetic field directly. This magnetic
field is extremely sensitive to
shear and compression motions, which are not easily detected in 2D velocity
fields. Using polarized radio-continuum emission at 6~cm, Vollmer et al. (\cite{vol04}) 
could unambiguously identify the ram-pressure compressed
region in the Virgo cluster spiral galaxy NGC~4522.
We were able to reproduce the H{\sc i} distribution, velocity field, and
the distribution of polarized radio-continuum emission with our
model simulations (Vollmer et al. 2006).

In order to understand the behavior of the large-scale magnetic field and thus
the polarized radio-continuum emission during a ram-pressure stripping event,
Otmianowska-Mazur \& Vollmer (\cite{kom03}) solved the induction equation on the
velocity fields provided by a dynamical model. They showed that a prominent
maximum of polarized radio emission is observed during compression and
in a state where the galaxy has passed the cluster center, the ram-pressure decreases,
and the gas pushed by ram pressure falls back onto the galactic disk.
We applied the method described in Otmianowska-Mazur \& Vollmer (\cite{kom03})
to NGC~4654; i.e., we solve the induction equation on the velocity fields of the simulations 
of Vollmer (\cite{vol03}) and compare the results to our polarized radio-continuum observations.

The Effelsberg and VLA radio continuum observations are presented in Sect.~\ref{sec:observations}.
In Sect.~\ref{sec:model} we explain the 3D MHD model and how we calculated the
model of polarized radio-continuum emission maps. The results of our model calculations are
shown in Sect.~\ref{sec:results} followed by direct comparison between our observations
and simulations (Sect.~\ref{sec:comparison}). 
These results are discussed in Sect.~\ref{sec:discussion} and we
give our conclusions in Sect.~\ref{sec:conclusions}.

\section{Radio observations of NGC~4654 \label{sec:observations}}


Radio polarimetric observations of NGC~4654 were performed at 8.35~GHz
using the single-horn receiver in the secondary focus of the 100-m
Effelsberg radio telescope  \footnote{The
100-m telescope at Effelsberg is operated by the Max-Planck-Institut f\"ur
Radioastronomie (MPIfR) on behalf of the  Max-Planck-Gesellschaft.}.
The galaxy has been scanned alternatively in R.A. and Dec. Four data
channels were recorded: two total power channels (then averaged to a single
total power Stokes I channel) and two correlation channels containing the
Stokes Q and U parameters. We obtained 33 coverages.

The telescope pointing was checked at time intervals of about 1.5 hours by
making cross-scans of a nearby strong point source. The flux-density scale
was calibrated by mapping the polarized source 3C138 and computing
its total power flux densities of 2.49~Jy at 8.35~GHz
using the formulae of Baars et al. (\cite{baars}).
The data reduction was performed using the NOD2 data reduction package
(Haslam \cite{haslam}). The coverages were combined into
final I, U, and Q maps using the spatial frequency weighting method
(Emerson \& Gr\"ave \cite{emgra}).

The distributions of Stokes parameters were combined into final maps of total
and polarized intensity, polarization degree, and polarization position
angle. Noisy structures smaller than the telescope beam-width were removed
using a digital filtering. A convolution to the beam-width of $90\arcsec$ was
applied to increase the sensitivity to the extended polarized emission.
The r.m.s. noise level in the final map of polarized
intensity at 8.35~GHz is 0.039~mJy/b.a.
\begin{figure}[ht]
\centering
\includegraphics[width=8.6cm]{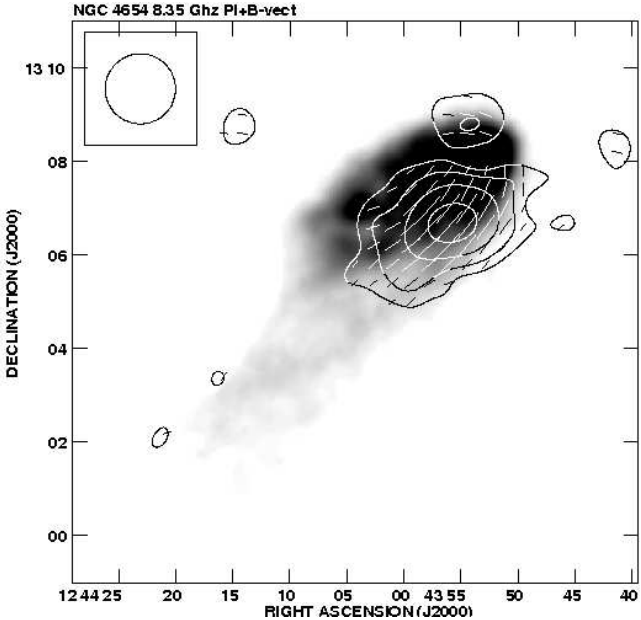}
\caption{Contours of polarized intensity of NGC~4654 at 8.35~GHz with
magnetic B-vectors proportional to the polarized intensity superimposed
onto the H{\sc i} emission. The contour levels are (3, 5, 8, 12) $\times$
0.039~mJy/b.a. The map resolution is 90\arcsec.
Greyscale: H{\sc i} gas distribution (Phookun \& Mundy \cite{pm95}).
}
\label{radioobs}
\end{figure}

NGC~4654 was observed, together with other large spirals of the Virgo Cluster,
and a detailed analysis of their radio total and polarimetric properties will
be discussed elsewhere. We also observed those galaxies at 4.85~GHz with
a resolution of 2.5\arcmin. The contour map of polarized intensity of
NGC~4654 at 8.35~GHz is presented in Fig.~\ref{radioobs} with
superimposed B-vectors of polarized intensity (i.e. E-vectors rotated by
90\degr), overlaid upon the H{\sc i} emission. The polarized emission shows strong
North-South asymmetry that may impose a compression or stretching of magnetic
fields on the southern side of the galaxy. Similar polarized structure is also
visible in our lower-resolution 4.85~GHz observations.


The VLA observations represent part of a larger
survey of 8 Virgo spiral galaxies that were observed in polarized radio
continuum emission at 6 and 20~cm (PI: B.~Vollmer).
Since allof the data will be presented in a forthcoming article,
we only show the polarized radio continuum emission at 6~cm.

\begin{figure}[ht]
\centering
\includegraphics[width=8.6cm]{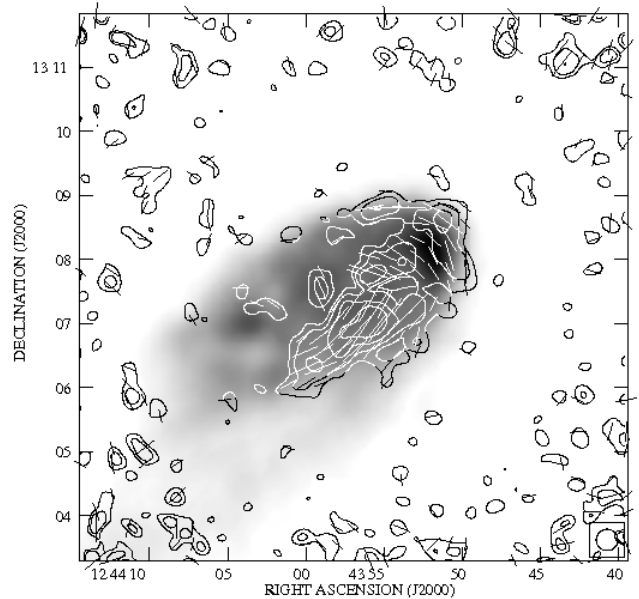}
\caption{Contours of polarized intensity of NGC~4654 at 4.85~GHz with
magnetic B-vectors proportional to the polarized intensity superimposed
onto the H{\sc i} emission. The contour levels are (3, 5, 8, 12, 20) $\times$
10~$\mu$Jy/b.a. The map resolution is 30\arcsec.
Greyscale: H{\sc i} gas distribution (Phookun \& Mundy \cite{pm95}).
}
\label{radioobsvla}
\end{figure}

NGC~4654 was observed at 4.85~GHz during 5~h on December 3 and 15, 2005 with the
Very Large Array (VLA) of the National Radio Astronomy Observatory
(NRAO)\footnote{NRAO is a facility of National Science Foundation
operated under cooperative agreement by Associated Universities, Inc.}
D array configuration. 
The band pass was $2\times 50$~MHz.
We used 3C286 as the flux calibrator and 1254+116
as phase calibrator, which was observed every 30~min. 
Maps of $17''$ resolution were made using the AIPS task 
IMAGR with ROBUST=3, which is close to pure natural weighting (5).
The maps were convolved to a beam size of the final maps of $20'' \times 20''$.
We ended up with an rms of 10~$\mu$Jy/beam in total and polarized emission. 
In Fig.~\ref{radioobsvla} the contour map of
polarized intensity of NGC~4654 at 4.85~GHz is presented, together with
superimposed B-vectors of polarized intensity (i.e. E-vectors rotated by
90\degr), overlaid upon the H{\sc i} emission.

Similar to the Effelsberg data, the VLA data show a prominent ridge
of polarized emission
in the south of the galaxy. Moreover, this ridge extends to the southwest.
The VLA data show that the lack of polarized emission in the Effelsberg data
at the tip of the H{\sc i} head is due to strong beam depolarization since
the magnetic field changes direction strongly following the H{\sc i}
and optical envelopes (Fig.~\ref{radioobsvla}). 
The magnetic vectors encircle the H{\sc i}
head-like structure in the northwestern part of the galaxy.

When we convolved the VLA data to the Effelsberg resolution, we found
two major differences between the VLA and Effelsberg maps:
(i) the northern maximum is more extended in the convolved VLA map,
which means that the Effelsberg telescope did not detect all the flux there
(the northern maximum is smaller than the Effelsberg beam; cf. Fig.~\ref{radioobs}), and
(ii) the southwestern part of the main ridge of polarized emission  in the Effelsberg data
seems to extend into the extended H{\sc i} tail. 
Since this is not observed in the VLA data, this additional
emission in the Effelsberg data is most probably due to the large beam.

Both features: the asymmetry in the polarized emission and the geometry of
the magnetic field in NGC~4654 should be explained by our MHD modeling and
can constitute a powerful tool for distinguishing between numerical experiments
and thus between different scenarios of gas dynamics (Vollmer 2003) 
and the magnetic field evolution in this galaxy.

\section{Model \label{sec:model}}

The 3D MHD model of magnetic-field evolution (see Otmianowska-Mazur \& Vollmer 
\cite{kom03} for more details) is applied to three selected numerical experiments of 
the Virgo galaxy NGC~4654 presented in Vollmer (\cite{vol03}). We take
only those cases into consideration where stars are treated as a
non-collisional component (Vollmer \cite{vol03}):
\begin{itemize}
\item[]{Model GR:} the gravitational (tidal) interaction with the second galaxy
\item[]{Model RPS:} the ICM ram pressure.
\item[]{Model GRPS:} the gravitational interaction and constant ICM ram pressure
\end{itemize}
The Zeus3D MHD code  (Stone \& Norman \cite{sn92}a and b) is used to solve
the induction equation
\begin{equation}
{\partial\vec{B}\over\partial t}=\nabla\times(\vec{v}\times\vec{B})
 -\nabla\times(\eta~\nabla\times\vec{B})
\label{eq:indeq}
\end{equation}
where $\vec{B}$ is the magnetic induction, $\vec{v}$ the large-scale
velocity of the gas, and $\eta$ the coefficient of a magnetic diffusion
(see Otmianowska-Mazur \& Vollmer~\cite{kom03}).
Realistic, time-dependent gas-velocity fields are
provided by the 3D N-body sticky-particle computations of H{\sc i} cloud
complexes evolving in  a galactic disk, a bulge and a halo simulated
by non-collisional particles (see Vollmer~\cite{vol03} and
Otmianowska-Mazur \& Vollmer~\cite{kom03}). The clouds 
collide inelastically, and in the two experiments, they are affected  
by the ICM ram pressure due to motion of the galaxy
with respect to the ICM. 

The velocity obtained from the N-body code has a discrete distribution.
In order to solve the induction equation (Eq.~\ref{eq:indeq})
with the help of Zeus3D code,
we have to interpolate these velocities to the regular grid. 
In Otmianowska-Mazur \& Vollmer (\cite{kom03}) we applied a spline
function with a density-dependent smoothing length. It turned out that we
had to use a large smoothing length  to suppress the noise 
in the velocity field of the outer disk, which was due to a small, 
local particle density.
In this way we avoided numerical artifacts of the magnetic field distribution
in the outer disk layers, but at the same time we obtained
a magnetic field distribution beyond the edges of the gas 
distribution.
For these reasons we decided to apply another method to
solve the interpolation problem.
The method is based on the kriging interpolation (see e.g Isaaks \&
Srivastava~\cite{krig89}) modified to be used effectively in our 3D case.

\subsection{Kriging interpolation and our modification}

Kriging is a family of linear least-square estimation algorithms. 
It is used to approximate or interpolate data.
The end result of kriging is to obtain the conditional expectation as a best estimate for all 
unsampled locations in a field and, consequently, a minimized error variance at each location. 
The kriging interpolation assumes (see e.g Isaaks \& Srivastava~\cite{krig89})
that the value of any (vector) field at a 
point $y(x)$  can be approximated by a weighted linear combination of a discrete set
of vectors given at a certain set of points $y_i=y(x_i)$ (provided by the N-body simulations, in
our case)
\begin{equation}
y(x)=\sum\limits_{i=1}^{N} y_i w_i(x)
\label{krigmain}
\end{equation}
where the weight coefficients $w_i(x)$ determine to what extent
$y_i$ influences $y(x)$.
The best approximation condition requires that the weights' sum has to be 
normalized by
$\sum\limits_{i=1}^{N} w_i(x)=1$. They can be determined from the
set of equations
\begin{equation}
\sum\limits_{i=1}^{N} d_{ij}w_i(x) + \lambda(x)=d_j(x)
\label{orgset}
\end{equation}
where $d_{ij}=d(x_i,x_j)$ and $d_i(x)=d(x_i,x)$ are called ``covariance
coefficients''. These coefficients describe how closeby the
points $x_i$ and $x_j$ (or $x_i$ and $x$) are distributed.
In our 3D case, we chose $d$ falling exponentially with the distance
$d_{ij}=e^{-||x_i-x_j||}$ and $d_i(x)=e^{-||x_i-x||}$.
A slack variable (the Lagrange coefficient)
$\lambda(x)$ is introduced here to balance numbers of unknowns with
$N+1$ equations ($N$ for each $w_i(x)$ plus the normalization equation).
This method implies determining (to solve a set of $N+1$ equations)
$w_i(x)$ and $\lambda(x)$ for each
point of interpolation $x$ -- in 171$\times$171$\times$71 points in our case.

We noticed that we can substantially reduce the calculation time.
If we define a new set of unknowns $g_i, i=0,1,\ldots,N-1$ and a
new slack variable $\kappa$ fulfilling the set of equations
\begin{eqnarray}
\label{g}    \sum\limits_{i=0}^{N-1} d_{ij}g_i + \kappa&=&y_j\\
\label{sumg} \sum\limits_{i=0}^{N-1} g_i&=&0,
\end{eqnarray}
then, multiplying equations (\ref{orgset}) by $g_j$ and summing over $j$, we get
\begin{eqnarray}
\sum\limits_{i=0}^{N-1} w_i(x) \sum\limits_{j=0}^{N-1} d_{ji} g_j
+ \lambda(x) \sum\limits_{j=0}^{N-1} g_j
&=&\sum\limits_{j=0}^{N-1} g_j d_j(x)
\end{eqnarray}
where we make use of the symmetry of $d_{ij}$. The second term vanishes
(Eq.~\ref{sumg}), and using Eq.~(\ref{g}), we can rewrite it as
\begin{eqnarray}
\sum\limits_{i=0}^{N-1} w_i(x) \left(y_i-\kappa\right)
&=&\sum\limits_{j=0}^{N-1} g_j d_j(x).
\end{eqnarray}
Finally, by substituting Eq.~(\ref{krigmain})
we obtain the following formula for the interpolated value
\begin{eqnarray}
y(x)&=&\sum\limits_{i=0}^{N-1} g_i d_i(x) + \kappa.
\label{fin}
\end{eqnarray} 
Thus, for all interpolation points we solve set of equations (\ref{g}) and
(\ref{sumg}) only once, and for each point $x$ we calculate values $d_i(x)$
and their scalar product with $g_i$.

We applied this interpolation routine only to the velocity-field components.
For interpolation of the mass distribution (used only for figures), we used a
much faster method -- smoothing particle masses with the Gaussian function
with smoothing-length small enough to see gaseous features.

\subsection{Model input parameters}

The dynamical models are fully described in Vollmer et al. (\cite{vol03}).
As in our previous paper (Otmianowska-Mazur \& Vollmer \cite{kom03}),
the induction equation is solved in rectangular coordinates ($XYZ$).
We use 171$\times$171$\times$71 grid points along the $X$, $Y$ and $Z$ axis, 
respectively. The grid spacing is 400~pc in all directions, which gives
the size of our modeled box as 
68.4\,kpc$\times$68.4\,kpc$\times$28.4\,kpc.  

We assume the magnetic field to be partially coupled to the gas via the 
turbulent diffusion process (Otmianowska-Mazur et al. \cite{kom00}) assuming
the magnetic diffusion coefficient to be $\eta=5\times10^{-25}$\,cm$^2$/s.
The initial magnetic field is purely toroidal with a strength of 10~$\mu$G
at a radius 2~kpc and falling linearly to zero toward $R=0$.
The magnetic-field strength is constant between $2$~kpc$< R < 20$~kpc
and decreases with a Gaussian profile for
$20$~kpc$< R < 30$~kpc with a half width of 500~pc.
It is zero for $R > 30$~kpc. In the vertical direction, the initial magnetic
field strength is a Gaussian function with a half width of 500~pc.

\subsection{The polarization maps \label{sec:polmaps}}

In order to obtain the simulated polarized intensity maps, we rotate
the cube of 3D magnetic field according to the orientation of  NGC~4654.
Next, we integrate the transfer equations of synchrotron emissivities in
Stokes I, Q, and U parameters along the line of sight: 
\begin{equation}
 {d \over d l}\left( \begin{array}[]{c} I\\ Q\\ U \end{array} \right)
  = \left( \begin{array}[]{ccc}
  \epsilon_I & 0 & 0\\
  p \epsilon_I \cos 2\chi & \cos\Delta & -\sin\Delta\\
  p \epsilon_I \sin 2\chi & \sin\Delta & \cos\Delta
  \end{array}\right)
  \left( \begin{array}[]{c} 1\\ Q\\ U \end{array} \right)
\label{eq:transfer}
\end{equation}
where the synchrotron emissivity is
\begin{equation}
 \epsilon_I\propto n_{rel}B_\perp^{(\gamma+1)/2}.
\end{equation}
We apply here the value of the relativistic electrons spectral index
$\gamma=2.8$ and the Faraday rotation angle 
\begin{equation}
 \Delta\propto \lambda^2  \int n_{th} B_\| {\rm d}s.
 \label{eq:faraday}
\end{equation}
The subscripts $\perp$ and $\|$ are used with respect to the line-of-sight,
and $\chi$ is the position angle of the sky-projected magnetic-field vector.
The integration in Eq.~\ref{eq:faraday} is along the line of sight,
and observations wavelength is $\lambda$.

The intrinsic degree of synchrotron polarization is assumed to be $p=75\%$.
Both thermal and relativistic electron distributions are taken as Gaussian
\begin{equation}
 n_{rel|th}\propto \mathrm e^{-(R/R_0)^2}\mathrm e^{-(z/z_0)^2} 
\label{eq:relel}
\end{equation}
where the radial scale-length $R_0$ is set to 10\,kpc and the
scale-height $z_0$ to 1\,kpc. We neglect Faraday effects by
setting $n_{th}=0$.
 
Finally for direct comparison with radio observations, the model maps of
Stokes I, Q, and U are convolved with a 2-D Gaussian 
function with a HPBW of 20$\arcsec$ and 80$\arcsec$. The final Q and U maps
are combined to obtain maps of polarized intensity and polarization
angle (rotated by 90$\degr$ to show the magnetic vector).

As described in Sect.~\ref{sec:introduction} the polarized radio-continuum
emission traces the ordered large-scale magnetic field. In general, the total magnetic
field of a spiral galaxy can be divided into two parts (see e.g. Beck et al. 1996): 
(i) a small-scale (smaller than a few hundred pc) magnetic field
that is connected to the turbulent motions of the interstellar medium and
(ii) a large-scale (larger than a few hundred pc)
ordered regular magnetic field that is most probably due to a large-scale $\alpha \Omega$-dynamo.
The large-scale magnetic field can be amplified 
by shear or compression motions of the interstellar medium where the magnetic field is
frozen in. The small-scale magnetic field is amplified by the turbulent
motions of the interstellar medium (small-scale dynamo).

The polarized radio-continuum emission is sensitive to a combination of 
magnetic-field compression or amplification. In addition, it is sensitive to the
average field ordering along the line of sight, which is a pure geometrical effect
depending on the position and inclination angle of the galaxy.
To separate these effects one can use the total power radio continuum map
to calculate the degree of polarization, which is only sensitive to the degree
of intrinsic field ordering.

As explained at the beginning of this section, we only model the large-scale
magnetic field (Eq.~\ref{eq:indeq}). We do not have access to the small-scale
magnetic field, which would require a model of the turbulent interstellar medium
together with a small-scale dynamo. Therefore, we are not able to calculate a 
total power map for the model. However, we plan for the future to tie the distribution
of the relativistic electrons (Eq.~\ref{eq:relel}) to the H$\alpha$ emission
distribution. This emission traces ionizing O and B stars that will
soon explode as supernovae giving rise to relativistic electrons.
If we assume that the star formation history is approximately constant
over the past ten million years, the H$\alpha$ emission represents
a reasonable measure for the relativistic electron distribution.
We can try to link the total power map to the H$\alpha$ emission using equipartition
between the turbulent kinetic energy of the interstellar medium (which is connected to the star 
formation rate and thus to the H$\alpha$ emission) and the total magnetic energy.

\section{Results  \label{sec:results}}

For the moment, we use the polarized radio-continuum emission for the
comparison between our models and observations. We show in the following that 
the information contained in the polarized radio-continuum is enough to
distinguish between the different dynamical models.

\begin{figure*}[ht]
\centering
\includegraphics[bb=10 40 435 435,width=0.29\textwidth]{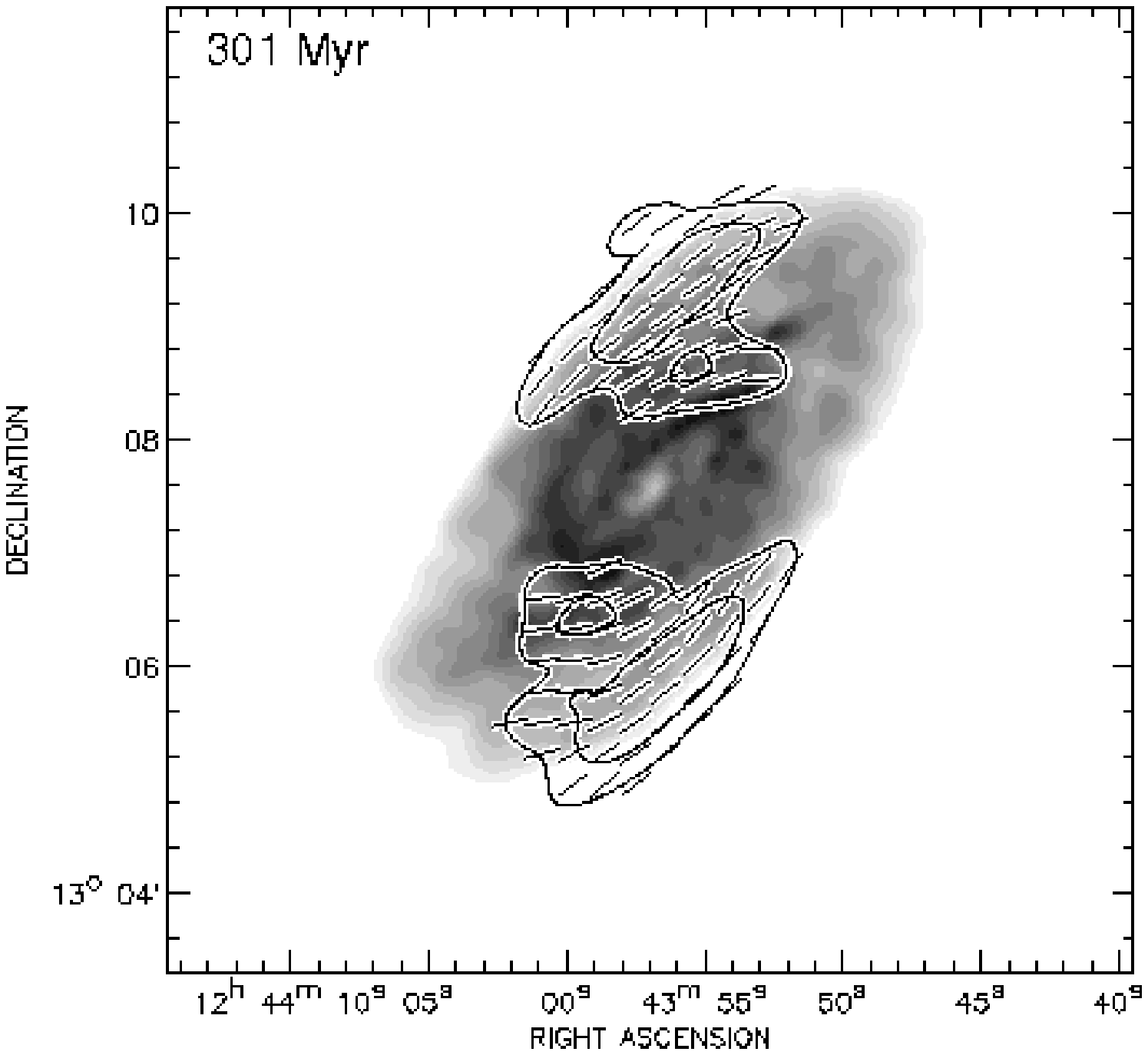}%
\includegraphics[bb=10 40 435 435,width=0.29\textwidth]{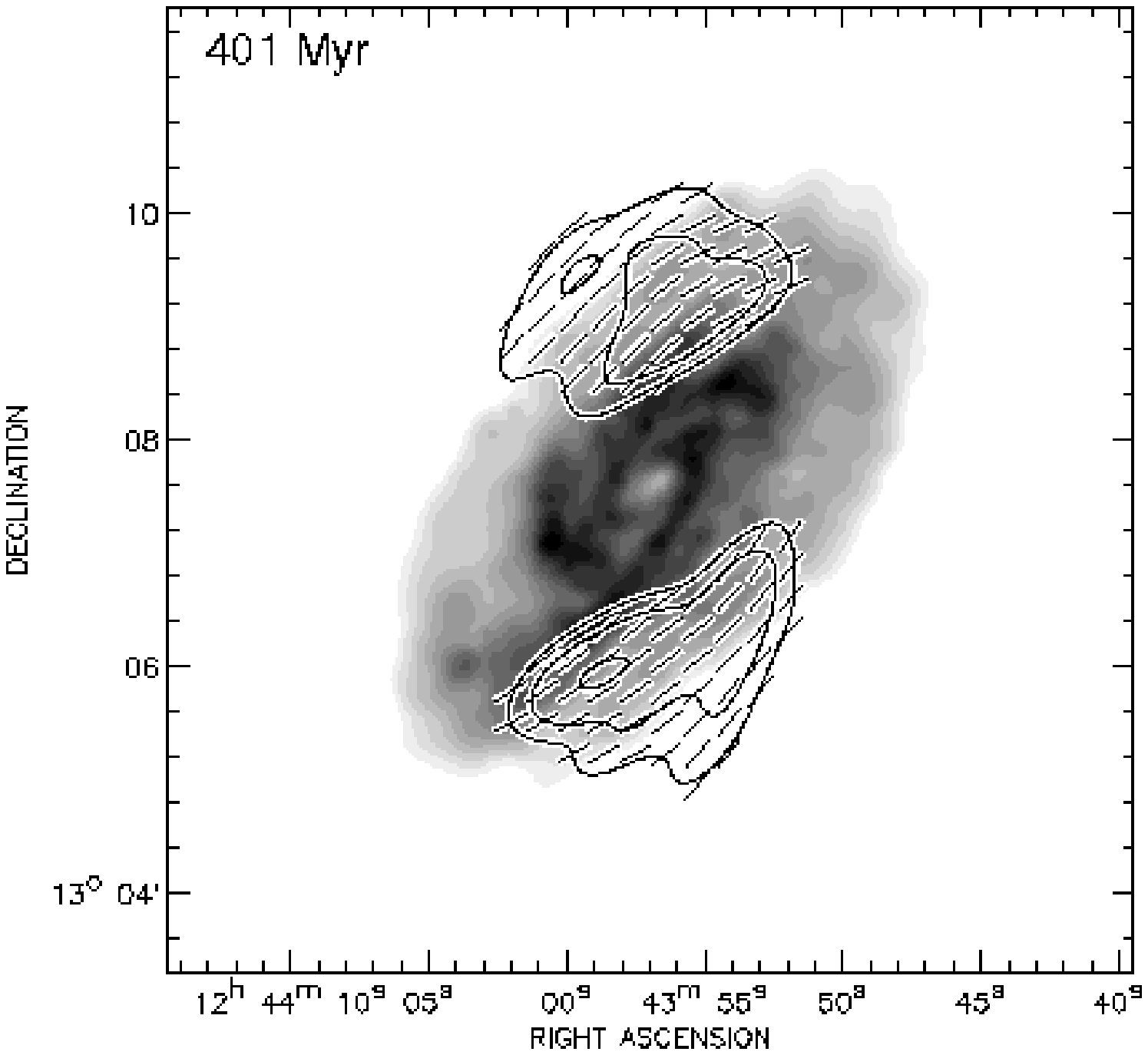}\\
\includegraphics[bb=10 40 435 435,width=0.29\textwidth]{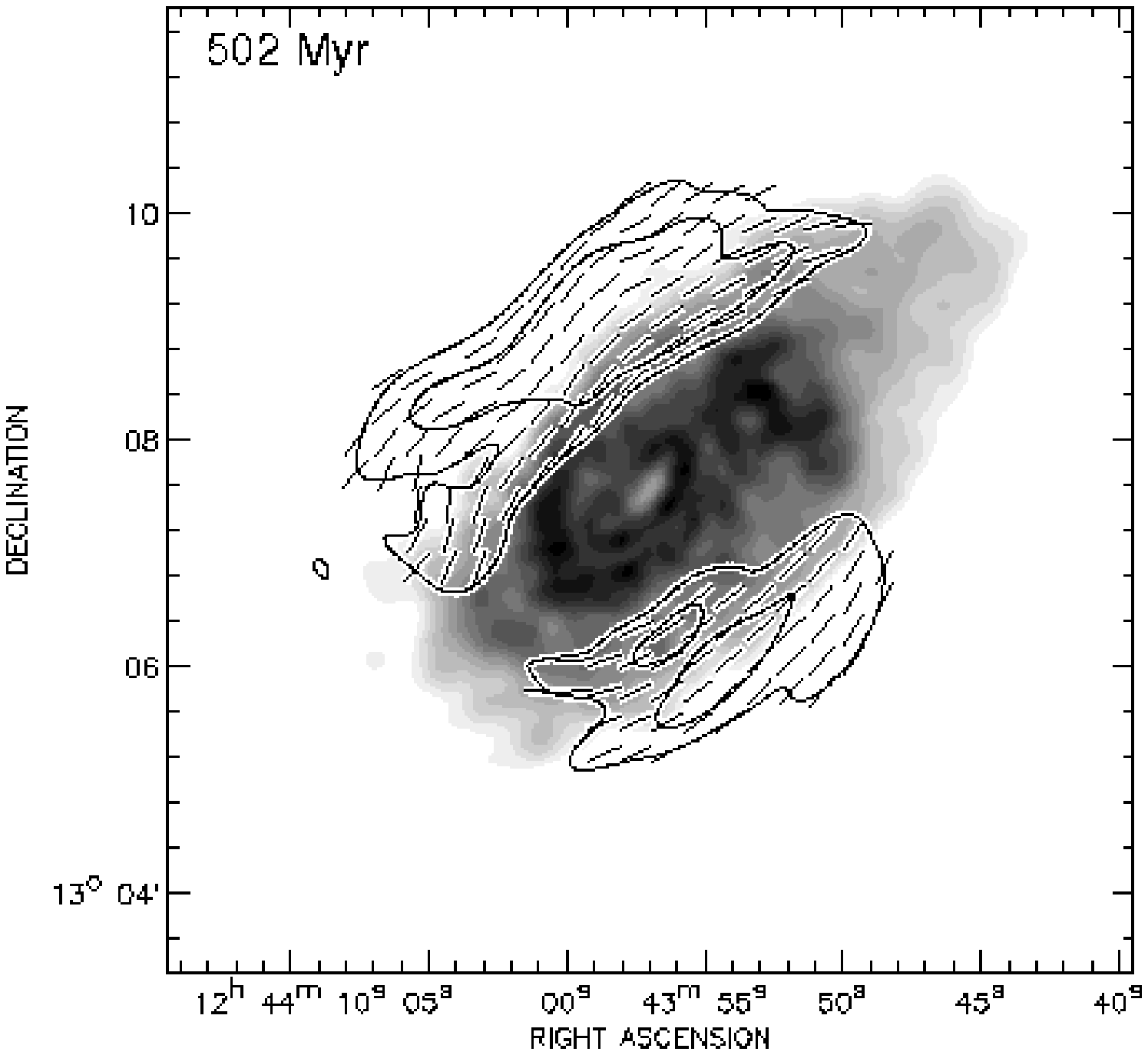}%
\includegraphics[bb=10 40 435 435,width=0.29\textwidth]{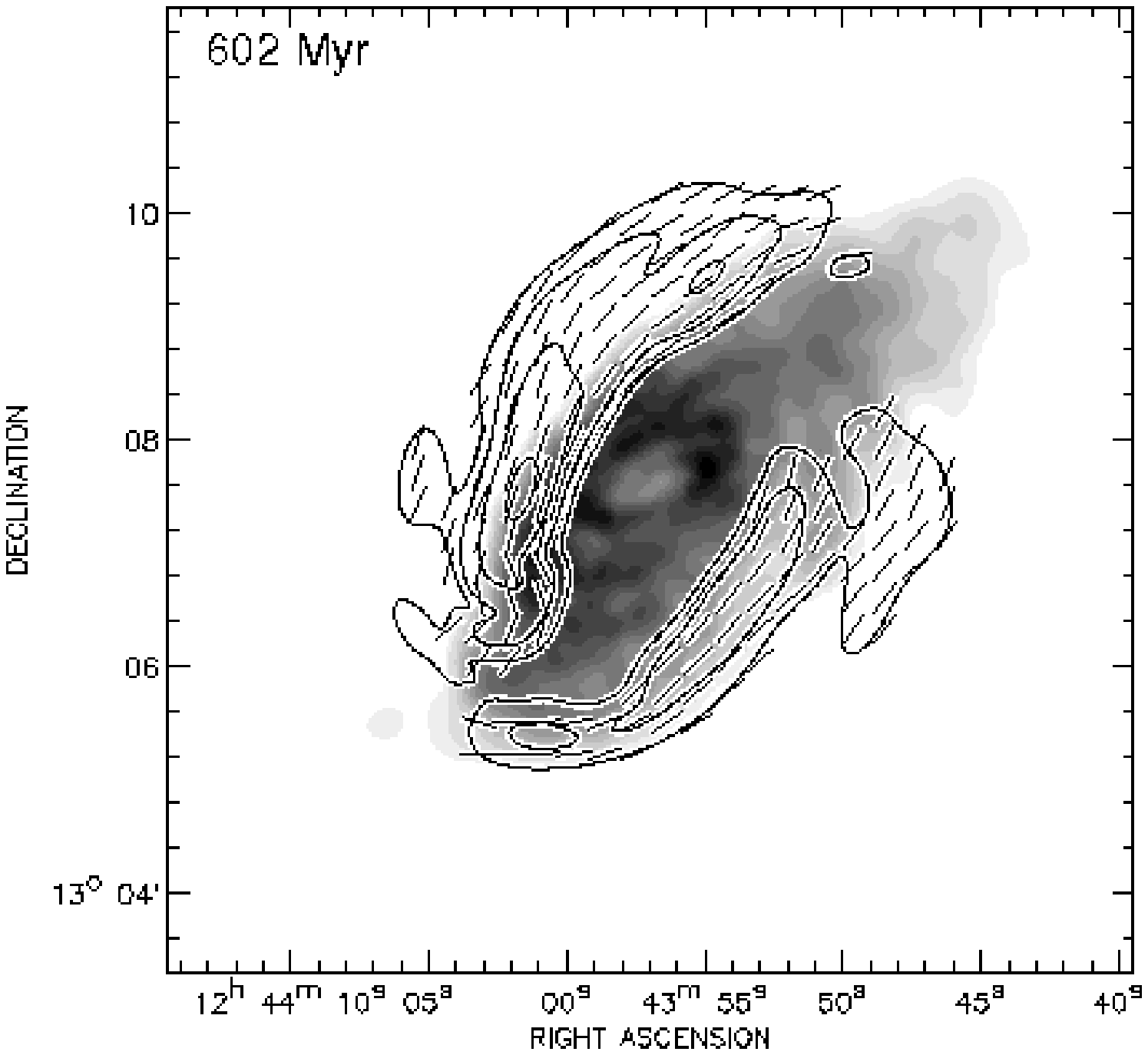}\\
\includegraphics[bb=10 40 435 435,width=0.29\textwidth]{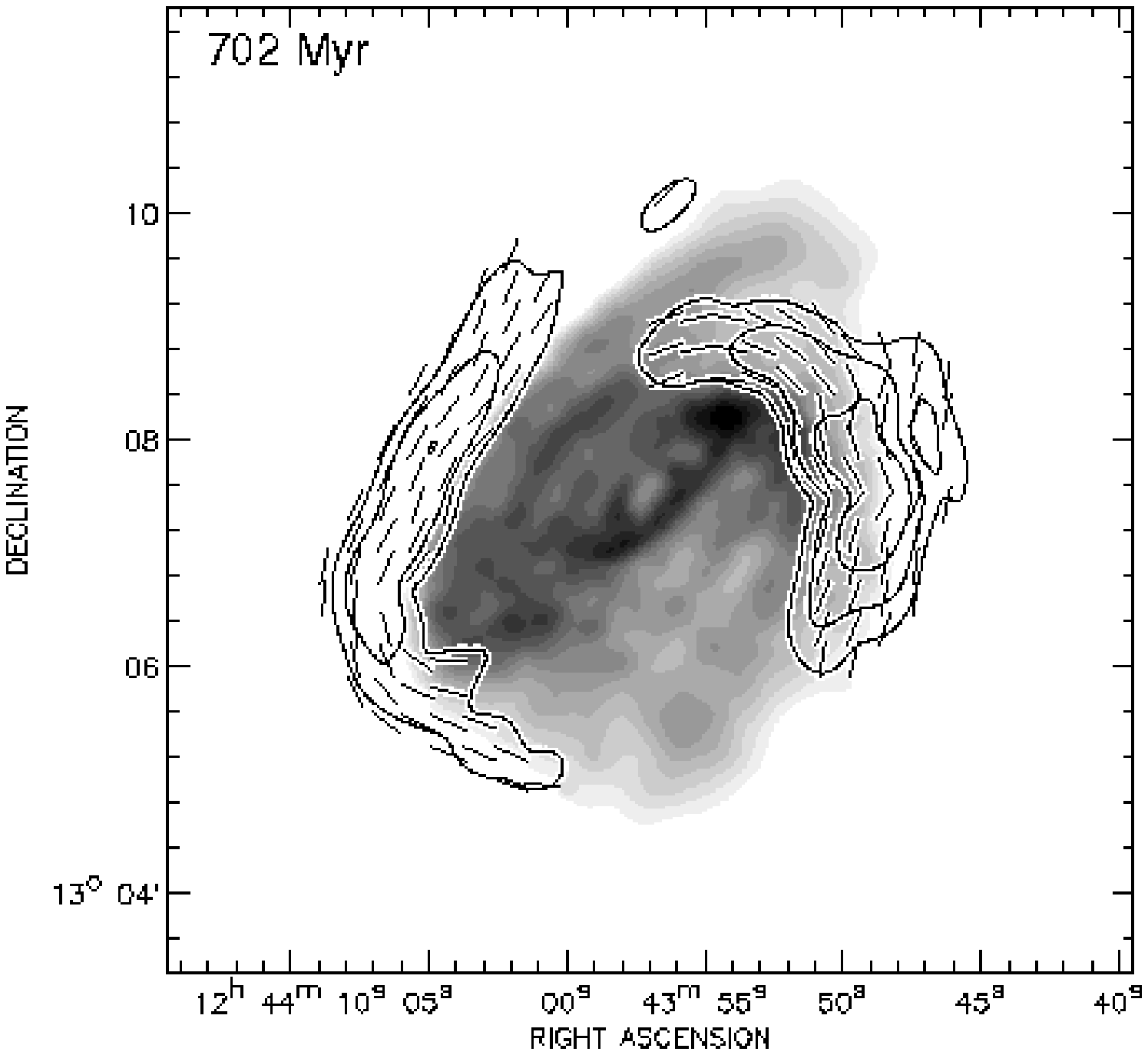}%
\includegraphics[bb=10 40 435 435,width=0.29\textwidth]{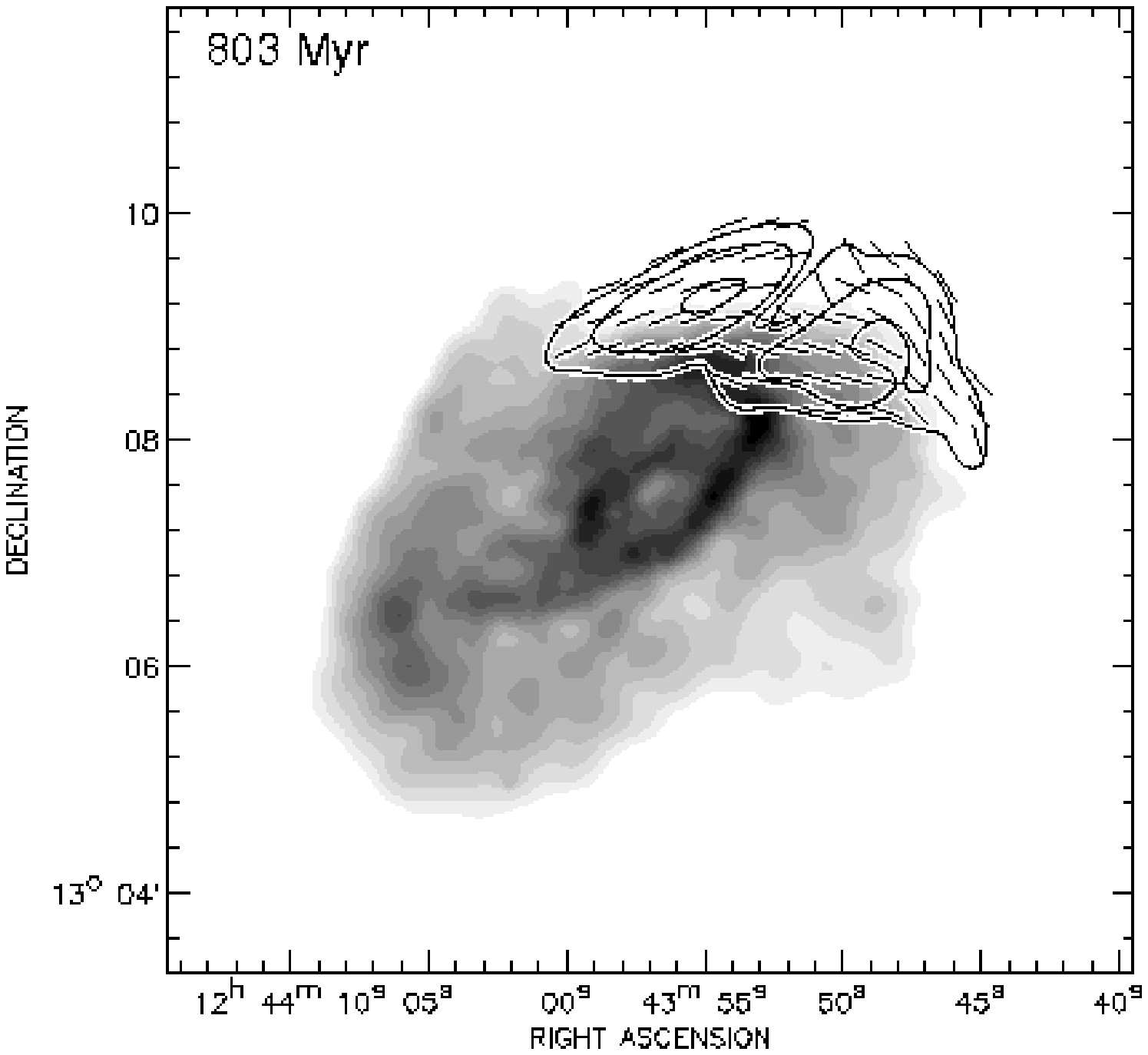}\\
\includegraphics[bb=10 40 435 435,width=0.29\textwidth]{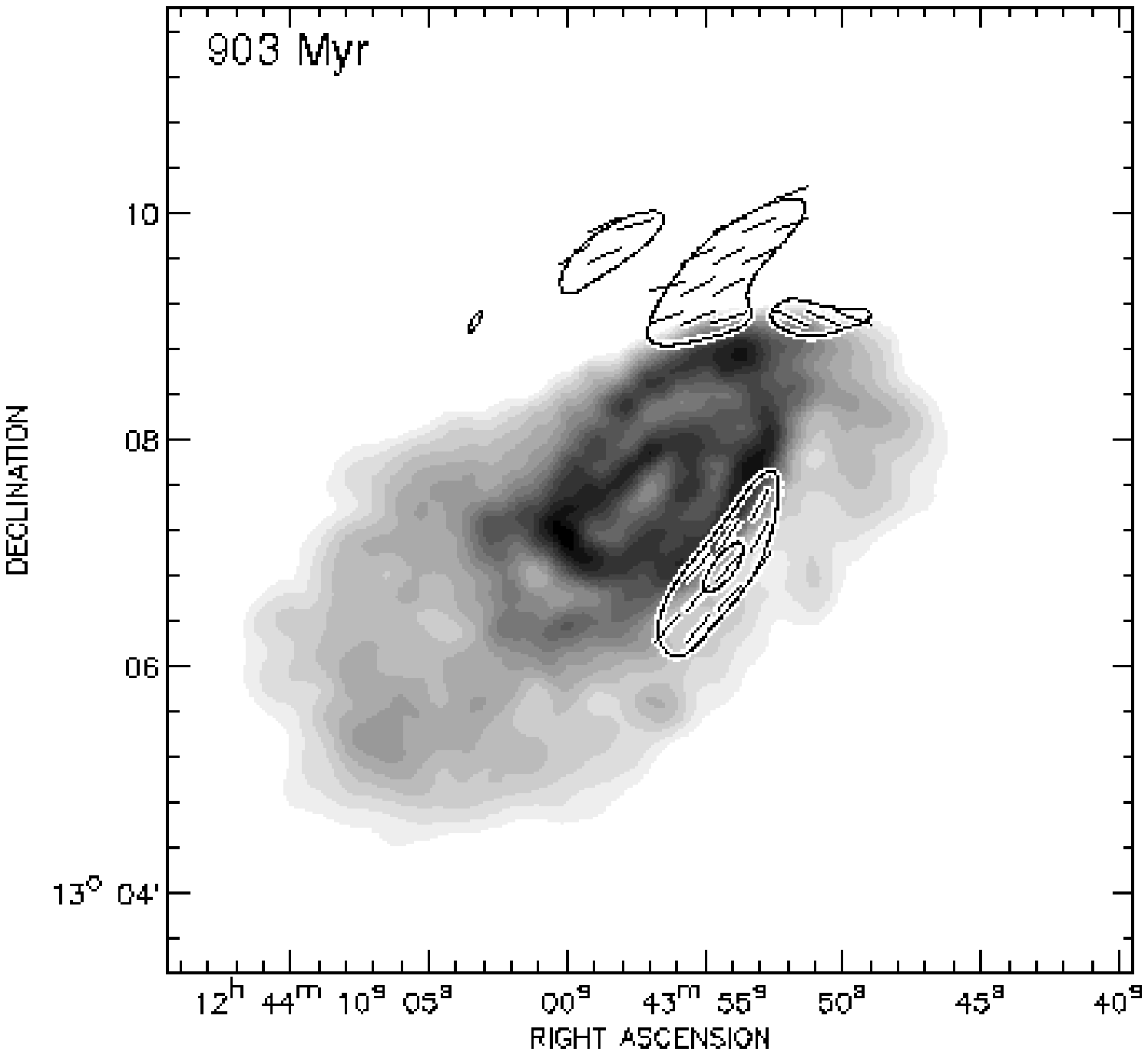}%
\includegraphics[bb=10 40 435 435,width=0.29\textwidth]{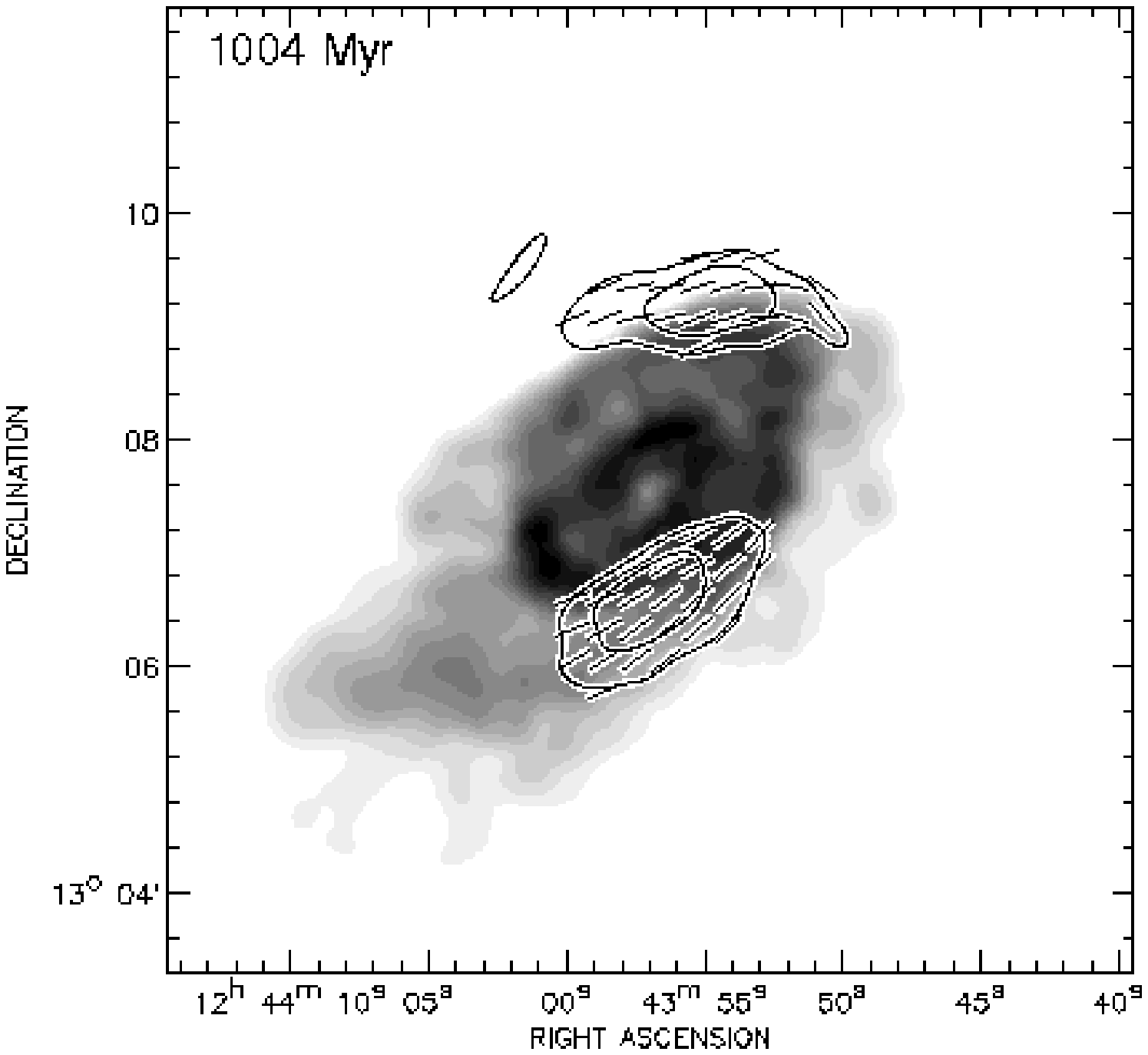}
\caption{
Evolution of the polarized intensity at chosen time steps
for the model only with gravitational interaction (model GR).
Contours: polarized intensity in logarithmic scale.
The polarized intensity
vectors are superimposed onto the gas density distribution in logarithmic scale
(grey plot).
The contours are plotted at levels of (1, 3, 10, 30, 100, 300)
(in arbitrary units) in all plots.}
\label{m106}
\end{figure*}

\subsection{Gravitational interaction alone (model GR)}

Figure~\ref{m106} shows the evolution of the polarized intensity distribution
for the model of tidally interacting galaxies.
The timestep of each snapshot is plotted in the upper left corner of each 
panel. The closest encounter between NGC~4654 and NGC~4639 occurs at
$t=460$~Myr. The images are convolved to a resolution of $20\arcsec$ HPBW.
This figure can be directly compared to Fig.~12 of Vollmer
(\cite{vol03}), because we present snapshots exactly at the same time steps.
Figure~12 of Vollmer (\cite{vol03}) also shows the positions of the 
interacting galaxy NGC~4639. We use the inclination and position angle
of NGC~4654 ($i=50$\degr{}, $PA=120$\degr{}).

The magnetic field distribution starts with an axisymmetric, purely azimuthal
configuration. Since the polarized radio-continuum emission is subject to 
Faraday rotation, i.e. the rotation of the magnetic field vector in the
plane of the sky due to the magnetic field component along the line of sight,
projection effects are important. The magnetic field component along the 
line of sight is largest at the two ends of the galaxy's major axis.
Therefore, Faraday depolarization suppresses the polarized radio-continuum emission there. 
Since the polarized radio-continuum emission depends on the 
line-of-sight integration of the
component of the magnetic field vector that lies in the plane of the sky,
geometrical effects are important. And since this component of the
magnetic-field vector is changing direction rapidly along the major axis,
the corresponding polarized radio-continuum emission is suppressed there.
As a result, the synchrotron emission distribution shows
two maxima to the northeast and southwest of the major axis.
The first snapshot in Fig.~\ref{m106} shows the polarized radio-continuum emission 
calculated using a slightly evolved magnetic field ($t=300$~Myr). 
During the galaxy evolution, non-azimuthal gas flows induce radial components of 
the magnetic field resulting in the non-zero magnetic pitch angles
without applying an explicit $\alpha$ dynamo mechanism. 
This effect becomes visible as a displacement of the polarized intensity maxima along the
major axis (see e.g. Knapik et al.~\cite{knapik}). 

At $t > 460$~Myr, the gravitational interaction with NGC~4639 perturbs the gas flow
leading to regions of gas compression and/or shear. Due to electromagnetic induction
(Eq.~\ref{eq:indeq}), the magnetic field is enhanced in these regions.
Thus, the polarized intensity maxima follow 
gas compression and shear regions, but because the timescale of turbulent diffusion 
is larger than the viscous timescale of the gas, there is no 
one-to-one correspondence between them. 
In some places and at some timesteps, gaseous and polarization features
coincide, while in other places and at other timesteps, arms of polarized
radio continuum emission lie just between gaseous ones (e.g. the northwestern
maximum at $t=$700~Myr).
More detailed analysis of polarized intensity distribution shows that in
general the maxima are left behind gas compression regions. This is not clearly 
visible in polarized intensity plots, as details are smoothed-out
due to the line-of-sight integration and beam convolution. 

Further evolution ($t > 800$~Myr) shows a weakening of the
magnetic field and thus a weakening of the polarized radio-continuum emission. 
In regions where gas falls back onto the galactic disk such as at the
basis of the southern tidal tail at $t \ge 900$~My,
the magnetic field is efficiently amplified, and maxima of polarized
emission reappear. In the final time step ($t=1000$~Myr), two comparable maxima of polarized 
radio continuum emission coincide with gas compression or shear regions.

\begin{figure*}[ht]
\centering
\includegraphics[bb=10 40 435 435,width=0.29\textwidth]{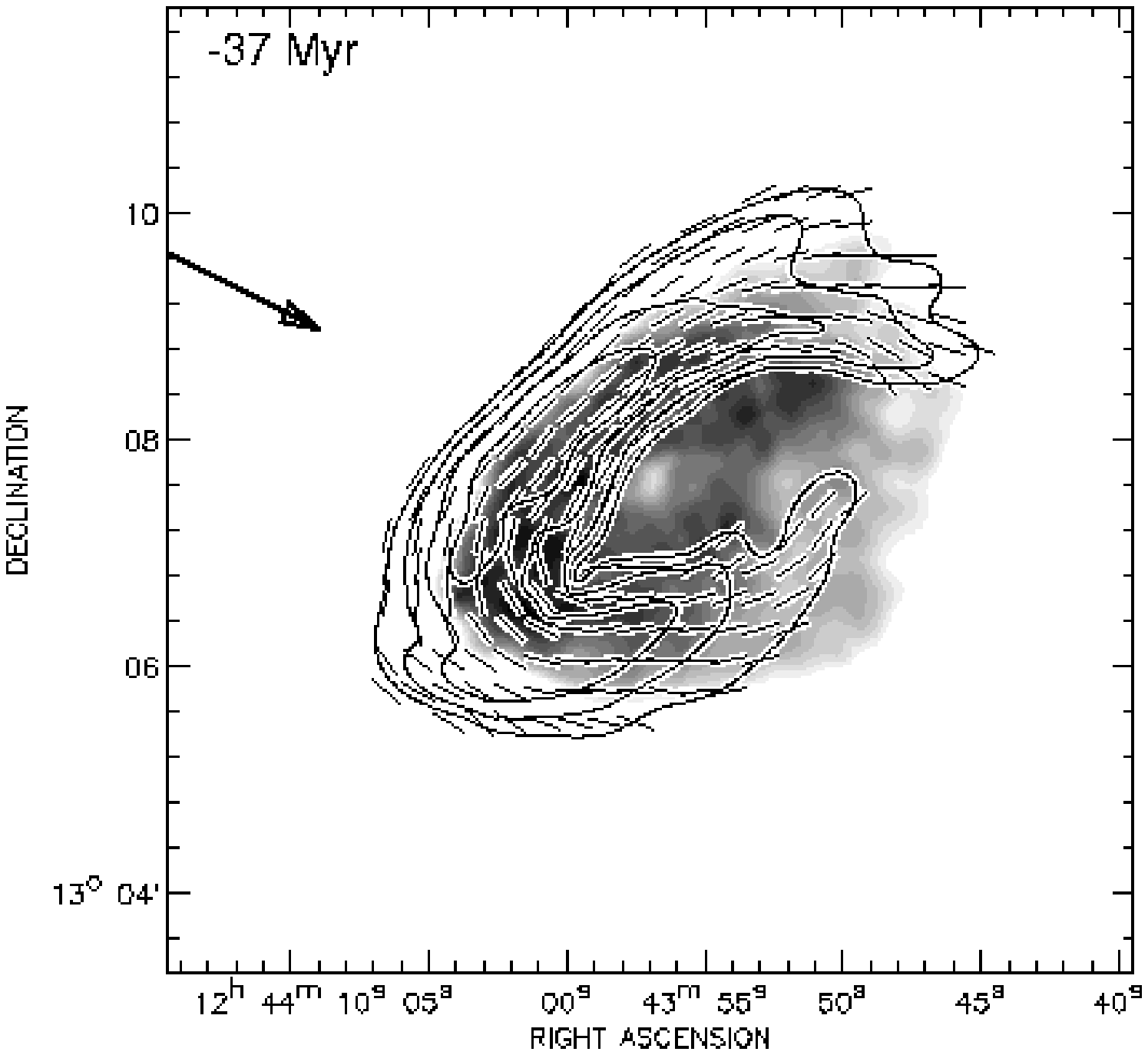}%
\includegraphics[bb=10 40 435 435,width=0.29\textwidth]{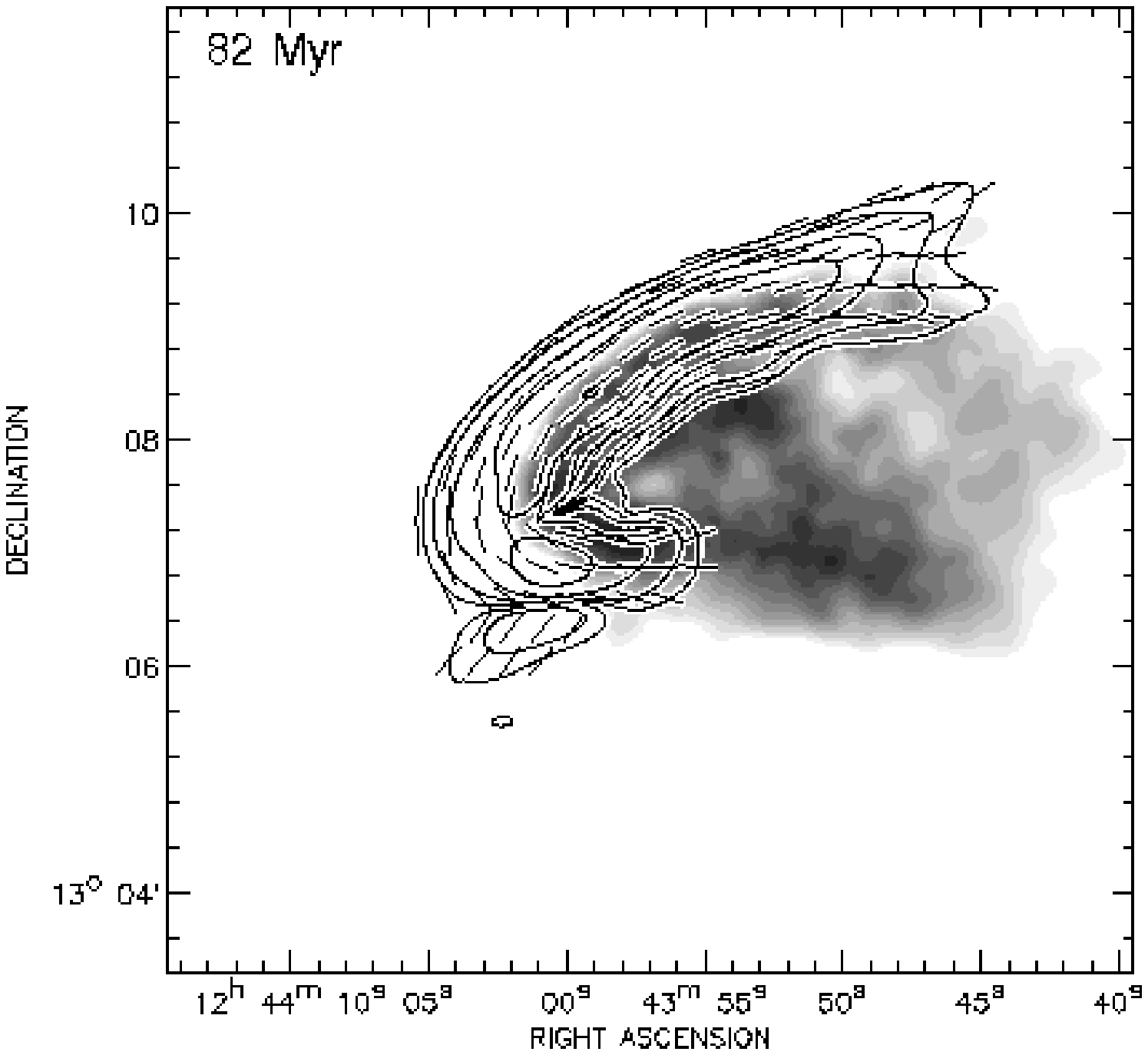}\\
\includegraphics[bb=10 40 435 435,width=0.29\textwidth]{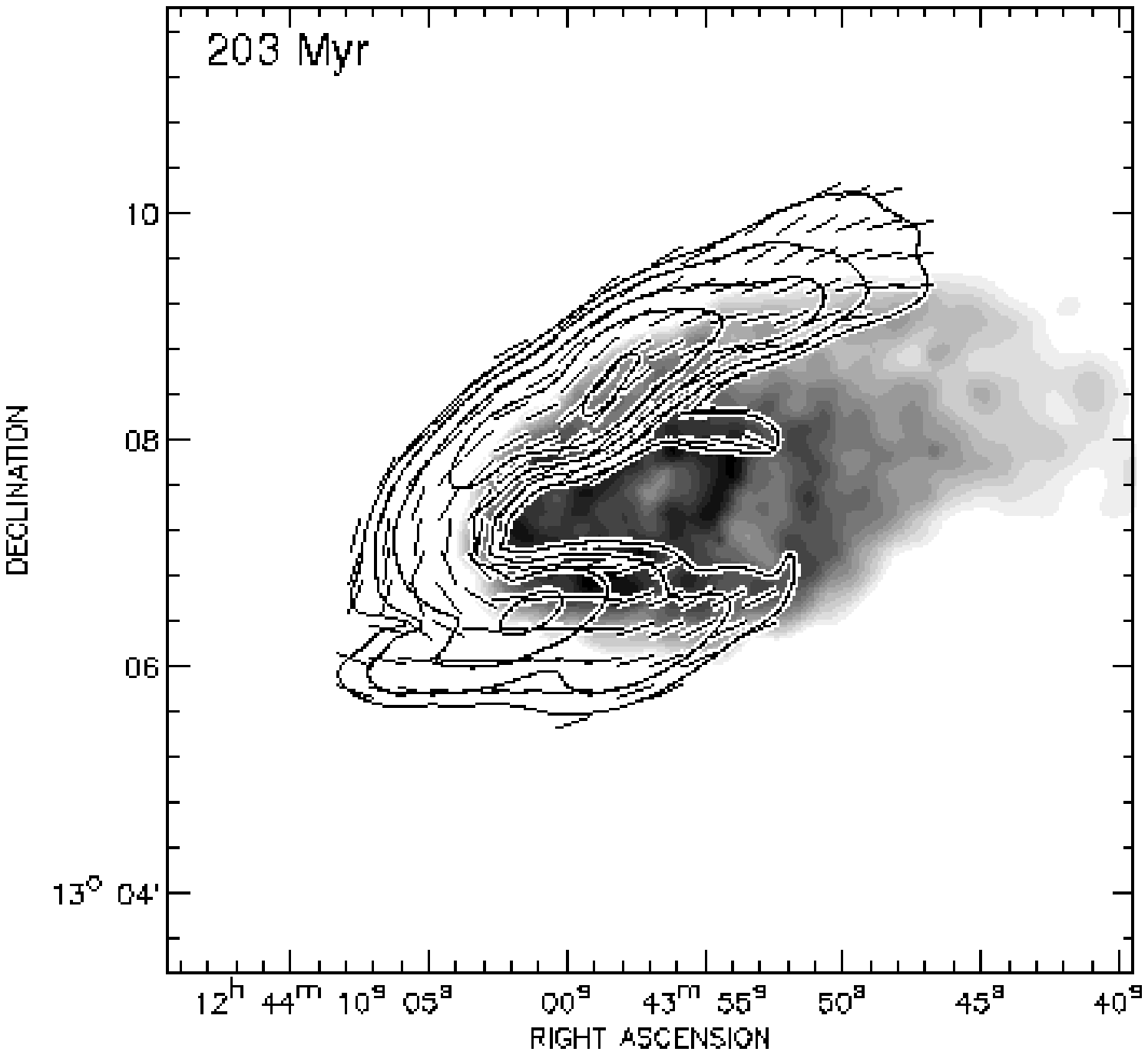}%
\includegraphics[bb=10 40 435 435,width=0.29\textwidth]{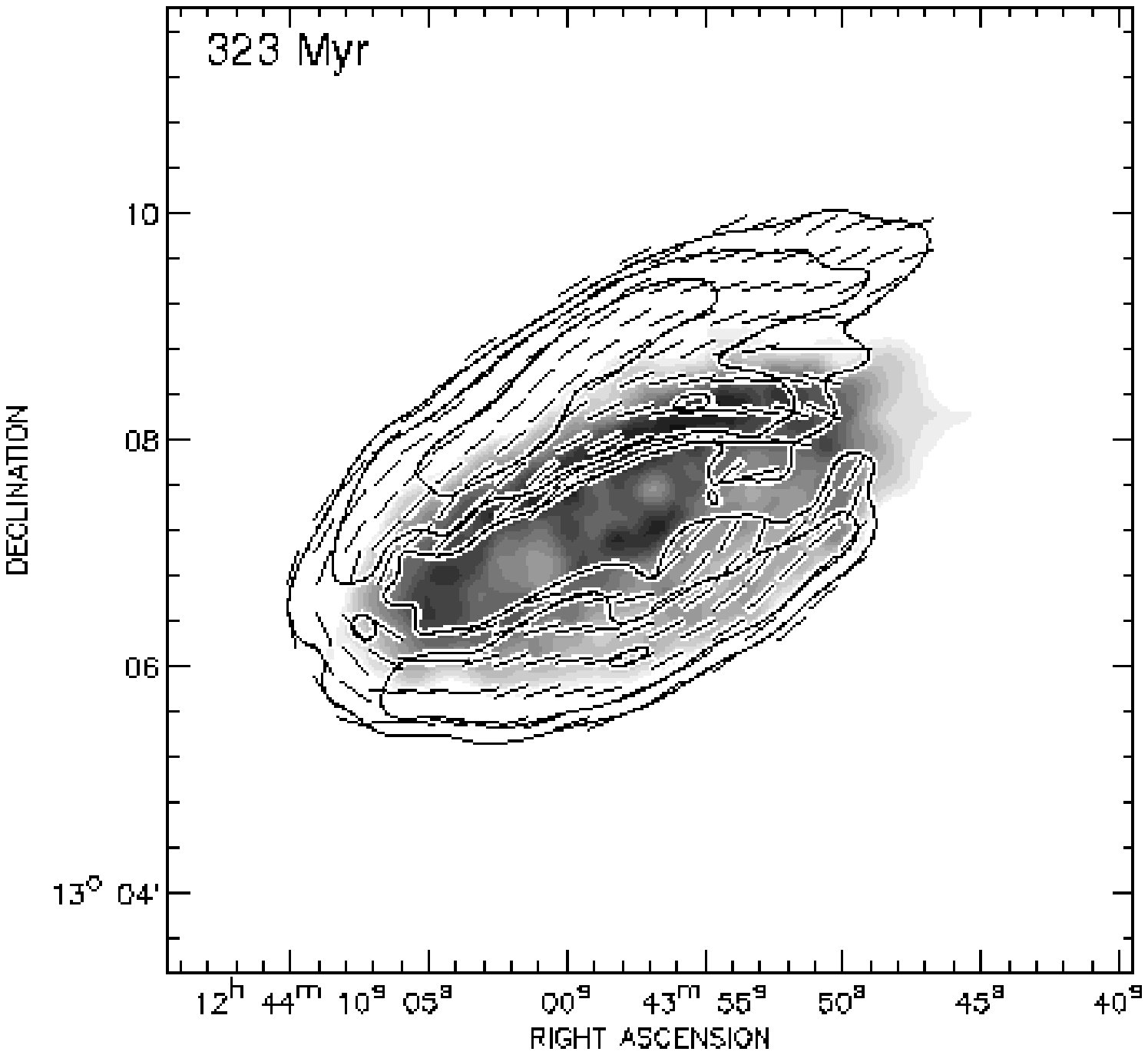}\\
\includegraphics[bb=10 40 435 435,width=0.29\textwidth]{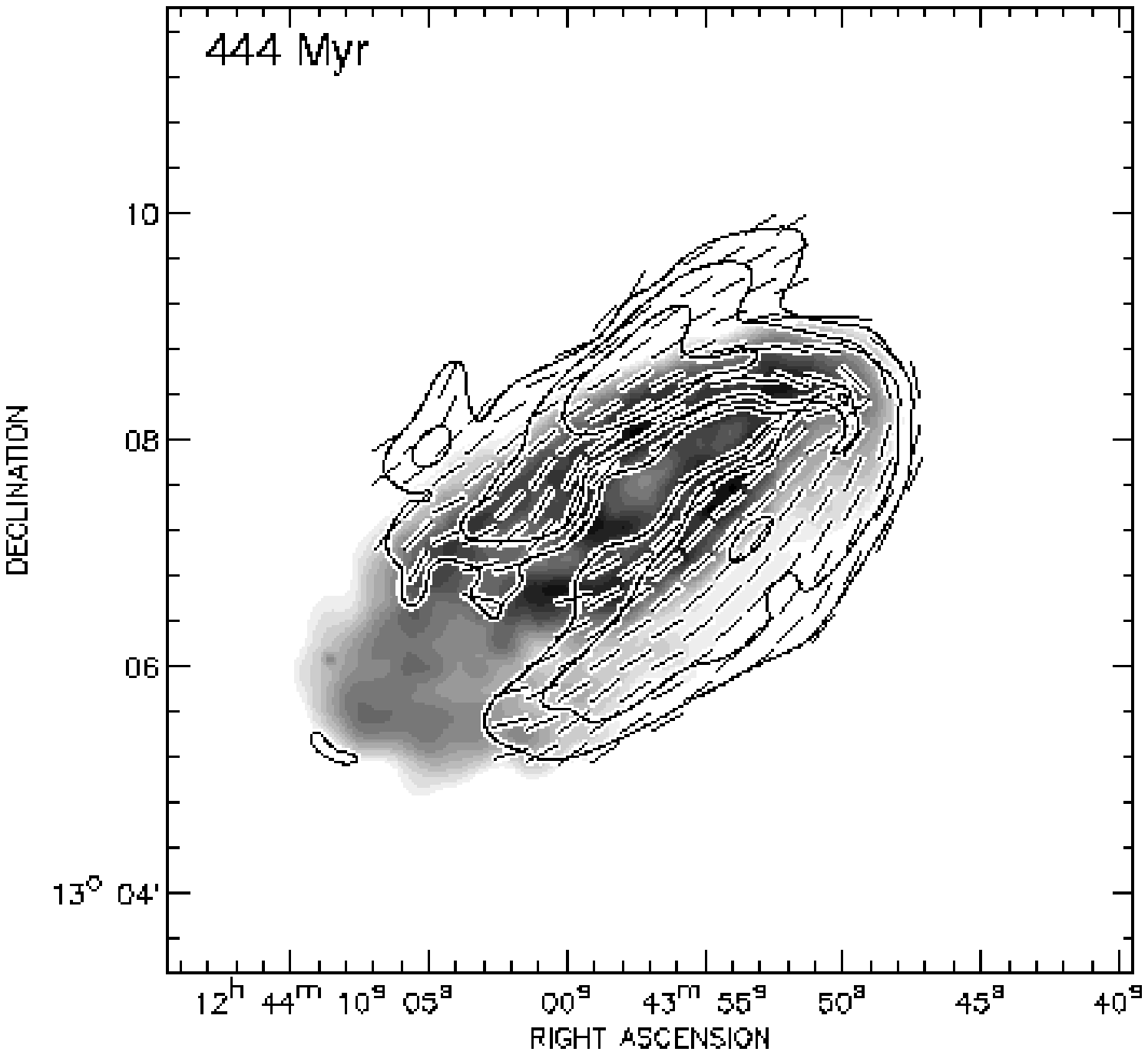}%
\includegraphics[bb=10 40 435 435,width=0.29\textwidth]{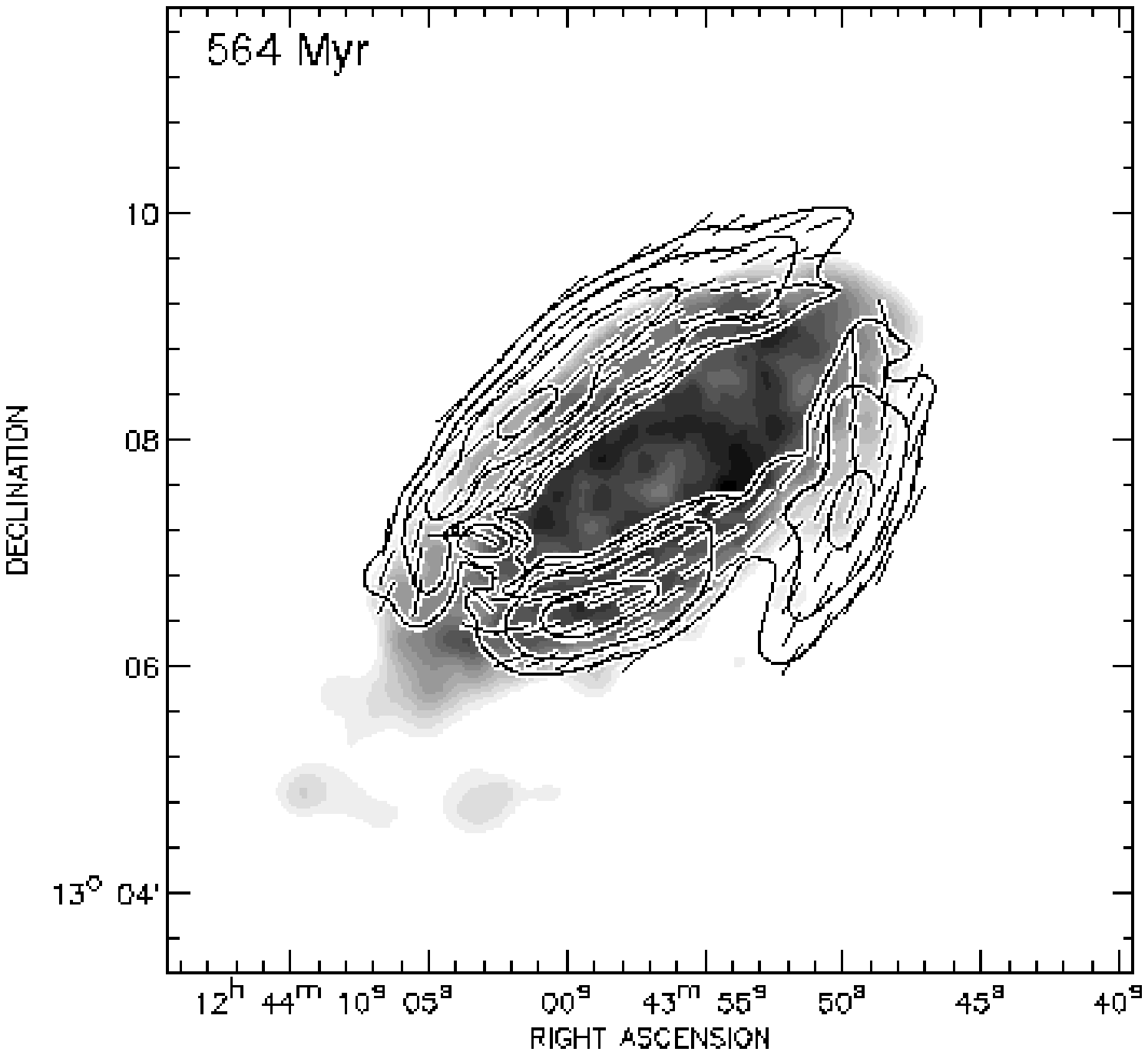}\\
\includegraphics[bb=10 40 435 435,width=0.29\textwidth]{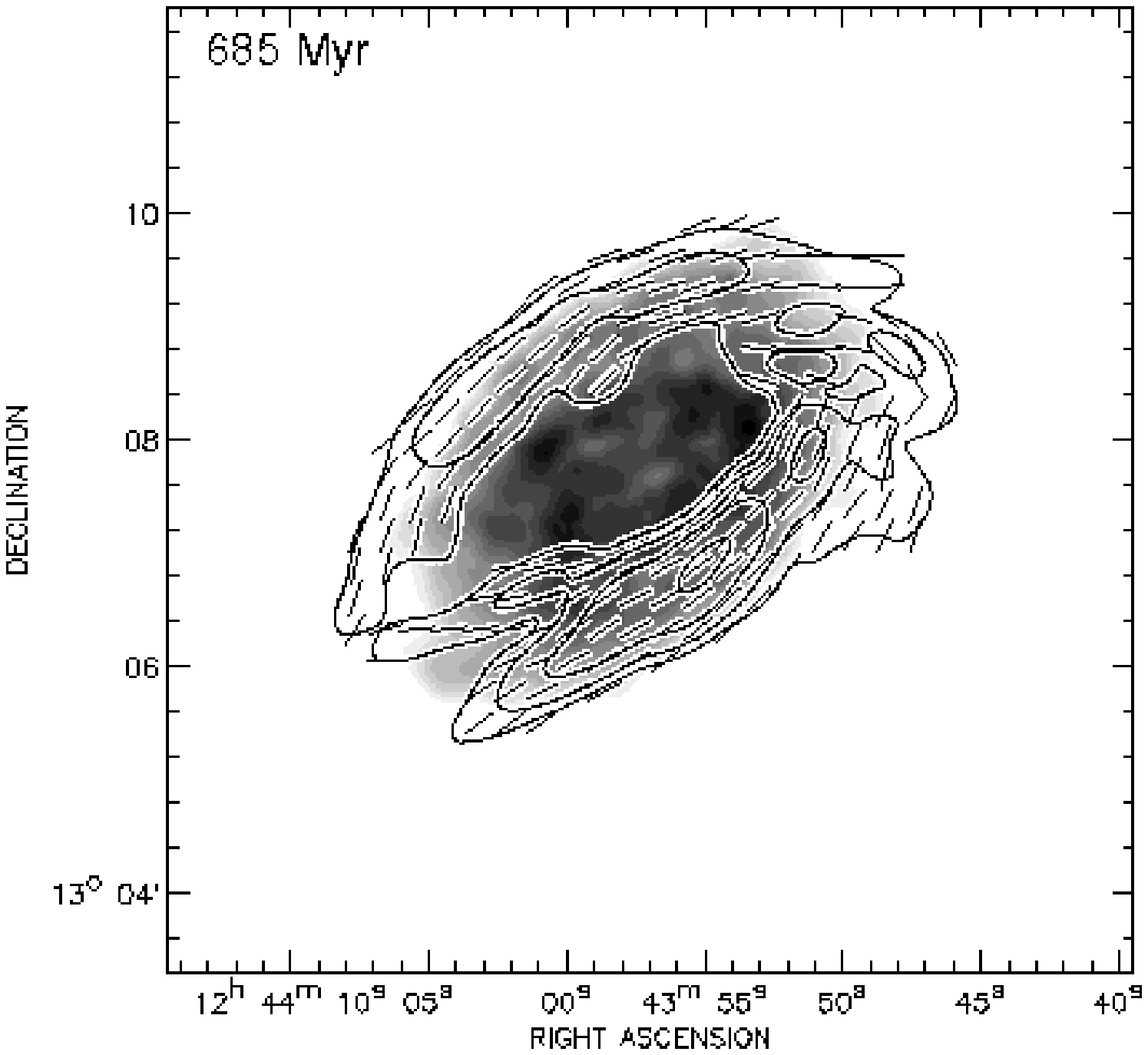}%
\includegraphics[bb=10 40 435 435,width=0.29\textwidth]{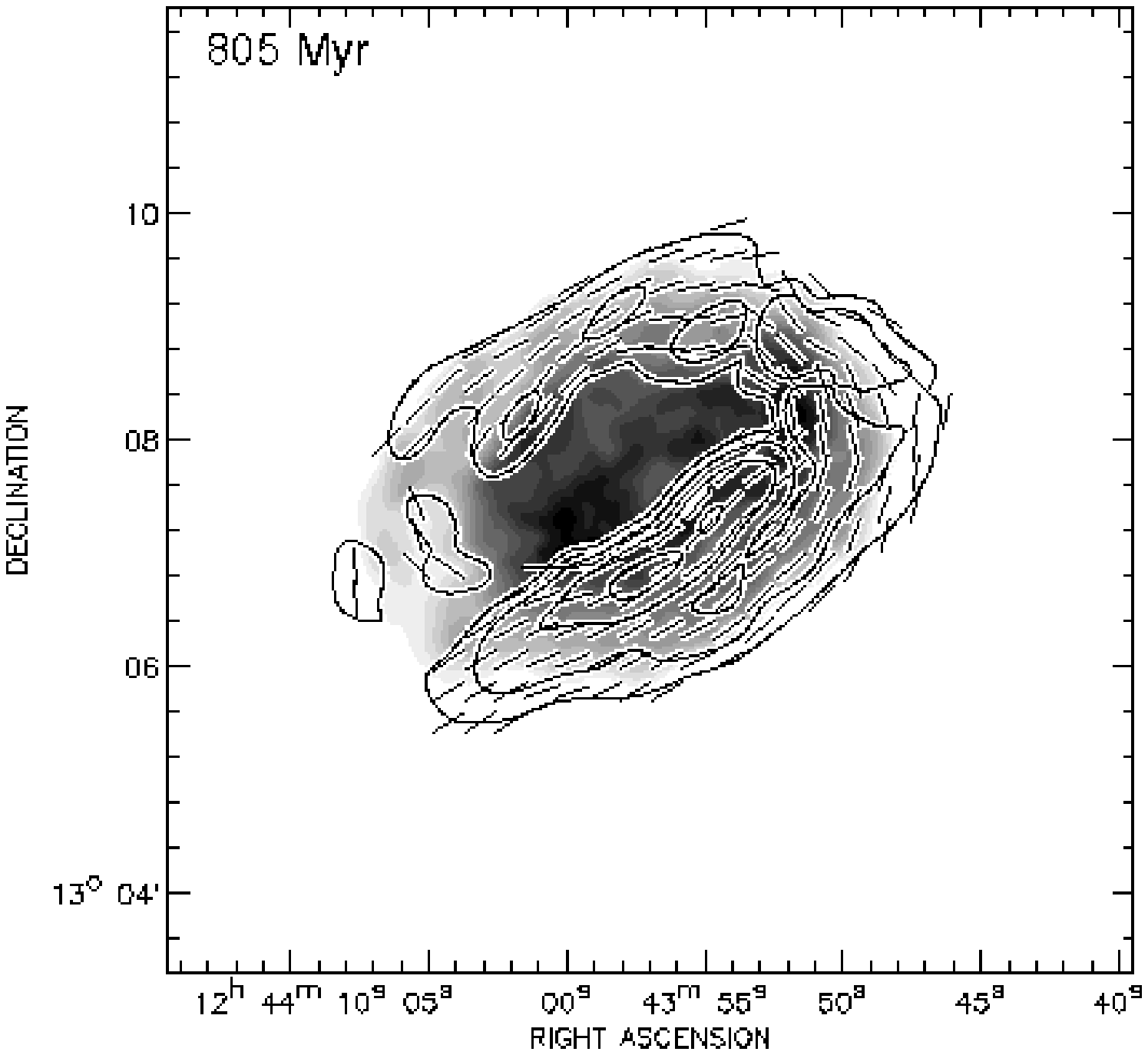}
\caption{
Evolution of the polarized intensity at chosen time steps
for the model with ram pressure event alone (model RPS).
Contours: polarized intensity in logarithmic scale.
The polarized intensity
vectors are superimposed onto the gas density distribution in logarithmic scale
(grey plot).
The contours are plotted at levels of (1, 3, 10, 30, 100, 300)
(in arbitrary units) in all plots.
The arrow in the top left panel indicates the direction of the
ram pressure wind, i.e. it is opposite to the galaxy's motion
within the intracluster medium.}
\label{m086}
\end{figure*}

\subsection{Ram pressure alone (model RPS)}

The second simulation only includes a strong ram-pressure stripping event.
The galaxy is diving deeply into the cluster potential where ram pressure
becomes stronger due to the increased velocity of the galaxy and the
higher intracluster medium density. The closest approach of the galaxy to the
cluster center (M87), and thus maximum ram pressure occurs at $t=0$~Myr.
The magnetic field evolution is strongly influenced by the gas compression
and shear motions induced by ram pressure (Fig.~\ref{m086}). 
This figure can be directly compared to Fig.~3 presented in
Vollmer (\cite{vol03}), because we present model snapshots at the same time steps.
The direction of the wind is identical to Fig.~3 of Vollmer (\cite{vol03}). 
The inclination angle between the galactic and the orbital plane is $i=0^{\circ}$.
The temporal ram pressure profile is a Lorentzian with a maximum ram pressure
of $p_{\rm ram}=5000$~cm$^{-3}$(km\,s$^{-1}$)$^{2}$.
In Otmianowska-Mazur \& Vollmer (\cite{kom03}) we calculated the evolution of
the distribution of polarized radio-continuum emission for a similar
ram-pressure stripping event where the inclination angle between the
galactic and the orbital plane was $i=20^{\circ}$ and maximum ram pressure
was $p_{\rm ram}=2000$~cm$^{-3}$(km\,s$^{-1}$)$^{2}$.
The evolution of the distribution of polarized radio-continuum emission
(Fig.~\ref{m086}) shows the same characteristics as that of 
Otmianowska-Mazur \& Vollmer (\cite{kom03}):
due to the strength of the interaction, the magnetic field distribution
and, in consequence, the maxima of polarized intensity follow almost perfectly
gas compression regions for $-50$~Myr$\le t \le 200$~My. 
Shortly before and after the maximum phase of the stripping event 
($t=-40$\,Myr and $t=80$\,Myr),
we observe a bright maximum of polarized radio-continuum emission 
in the northeast of the galaxy center, at the location where the galaxy's ISM 
is compressed by ram pressure. 
At $t=200$~Myr ram pressure and compression decrease and rotation takes
the magnetic field along with it. Since the galaxy rotates counter-clockwise,
the result is an additional maximum of polarized emission in the south of the
galaxy center. Rotation then makes the distribution of polarized emission
approximately circular ($t=320$~Myr). At $t > 500$~Myr the stripped gas that
was not accelerated to the escape velocity begins to fall back onto
the galactic disk in the southeast of the galaxy center (re-accretion).
This leads to shear motion that
enhances the large-scale magnetic field and thus the polarized radio-continuum
emission. Re-accretion of stripped gas then becomes stronger
with time leading to a stronger maximum of polarized emission in the south of
the galaxy center (see also Fig.~9 of Otmianowska-Mazur \& Vollmer \cite{kom03}). 
Differences between the polarized emission and the gas distributions are due 
to (i) the longer timescale of turbulent diffusion compared to the viscous
timescale and (ii)
orientation (integration along the line-of-sight) and beam depolarization
effects. The final stage of the magnetic field evolution shows an approximately
circular distribution of polarized radio-continuum emission with
a maximum in the north and another maximum to the south of the galaxy center. 
The southern maximum is brighter than the northern one.

\subsection{Gravitational interaction and ram pressure (model GRPS)}

\begin{figure*}[ht]
\centering
\includegraphics[bb=10 40 435 435,width=0.29\textwidth]{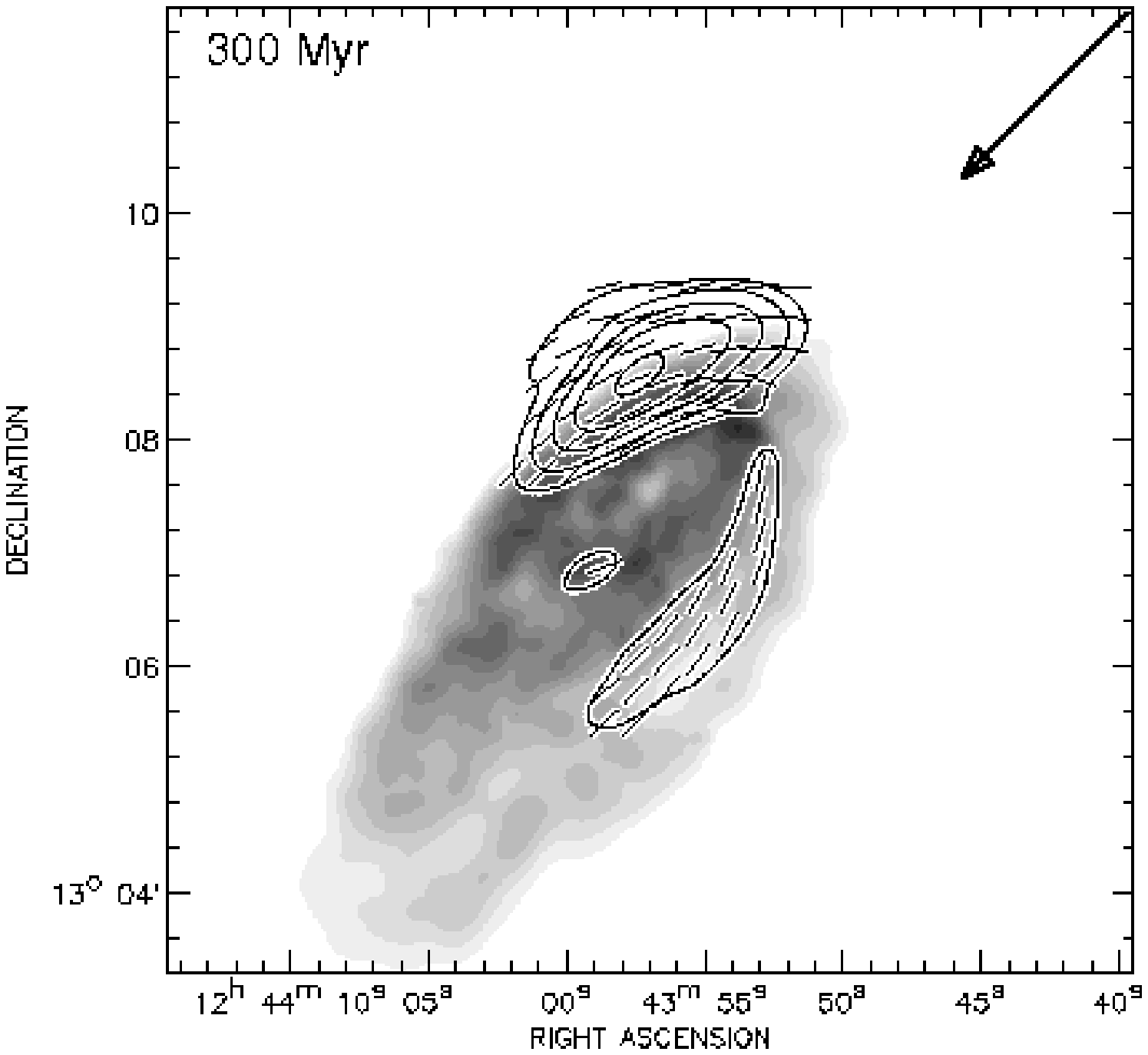}%
\includegraphics[bb=10 40 435 435,width=0.29\textwidth]{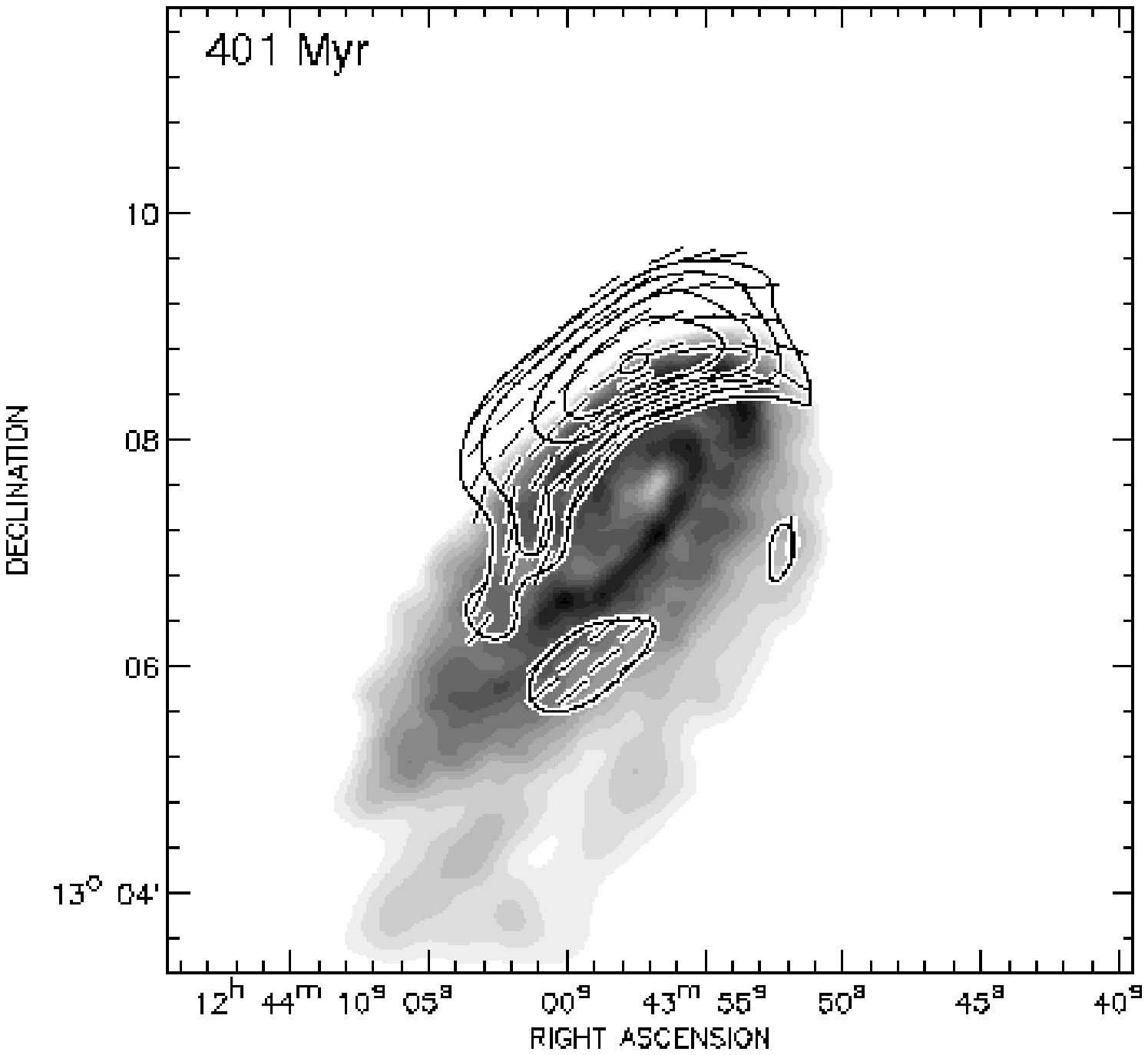}\\
\includegraphics[bb=10 40 435 435,width=0.29\textwidth]{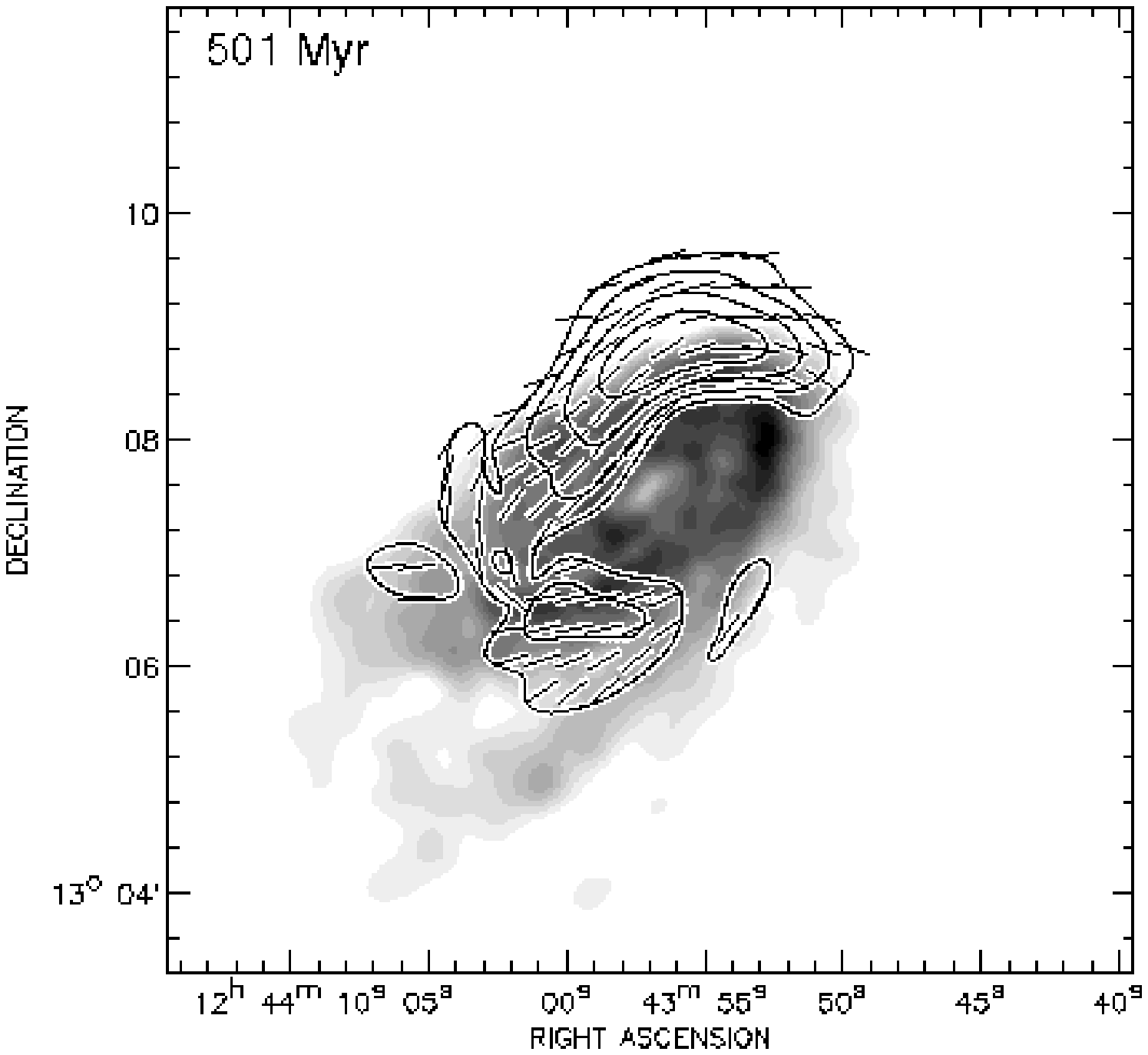}%
\includegraphics[bb=10 40 435 435,width=0.29\textwidth]{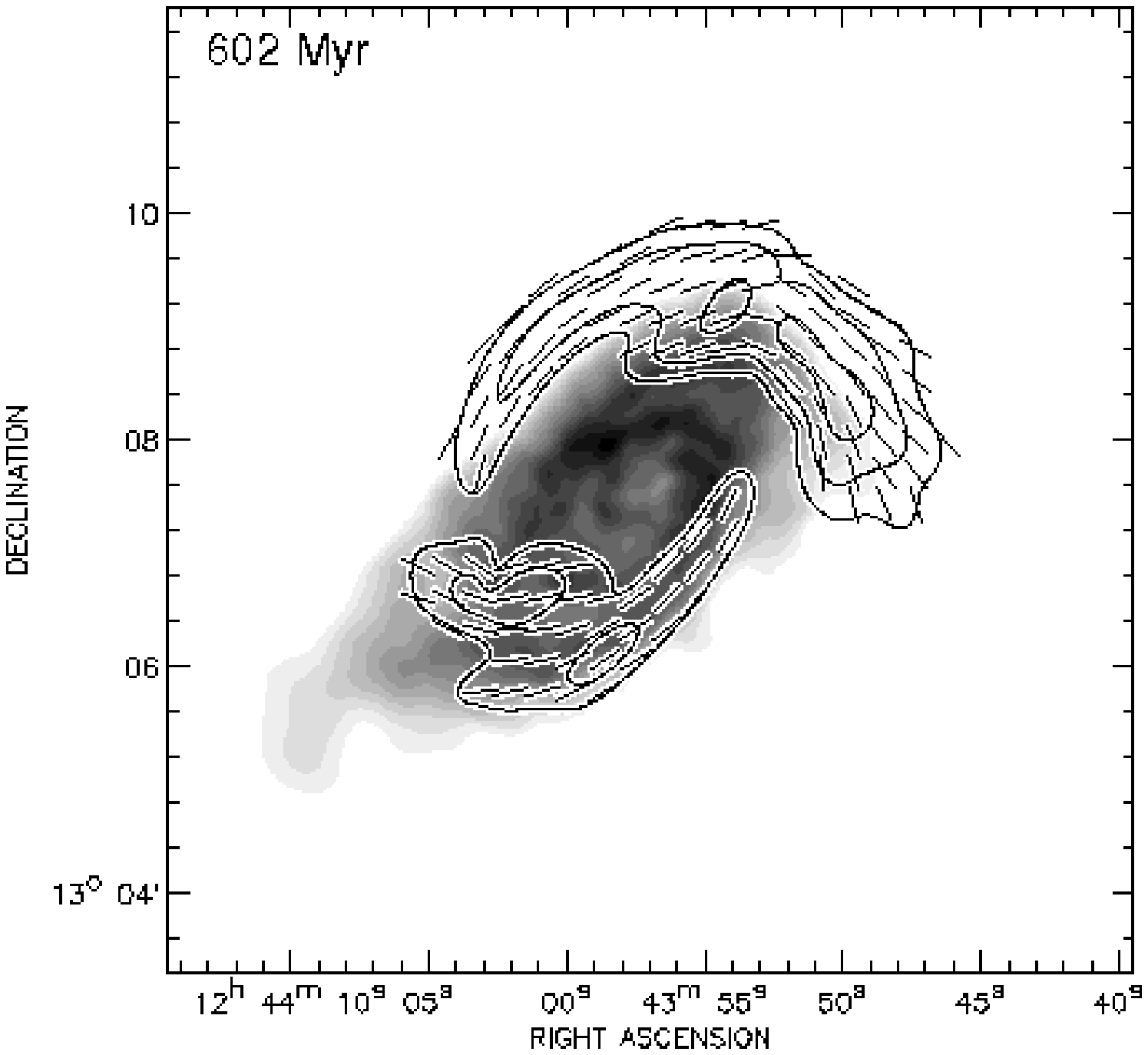}\\
\includegraphics[bb=10 40 435 435,width=0.29\textwidth]{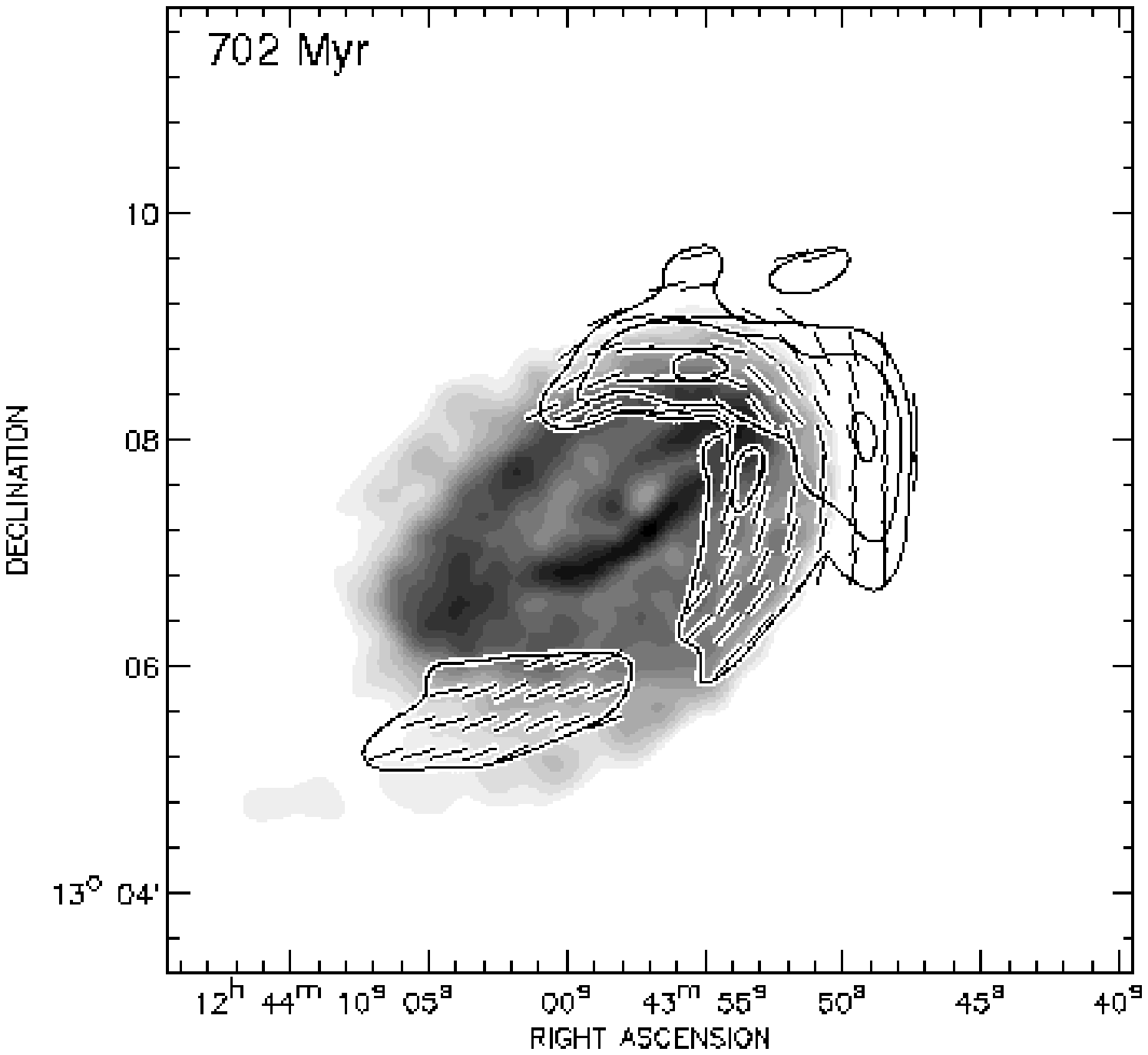}%
\includegraphics[bb=10 40 435 435,width=0.29\textwidth]{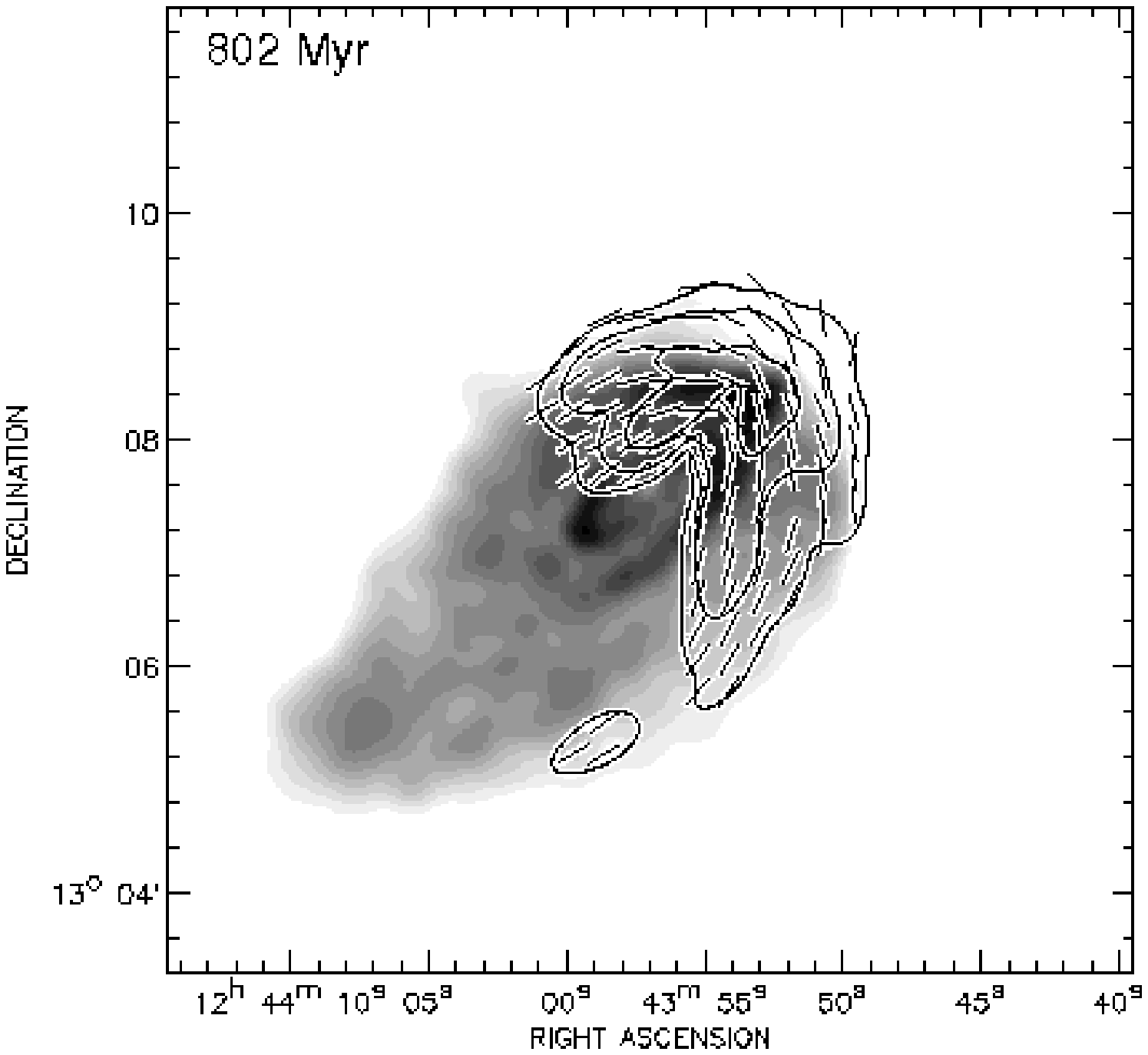}\\
\includegraphics[bb=10 40 435 435,width=0.29\textwidth]{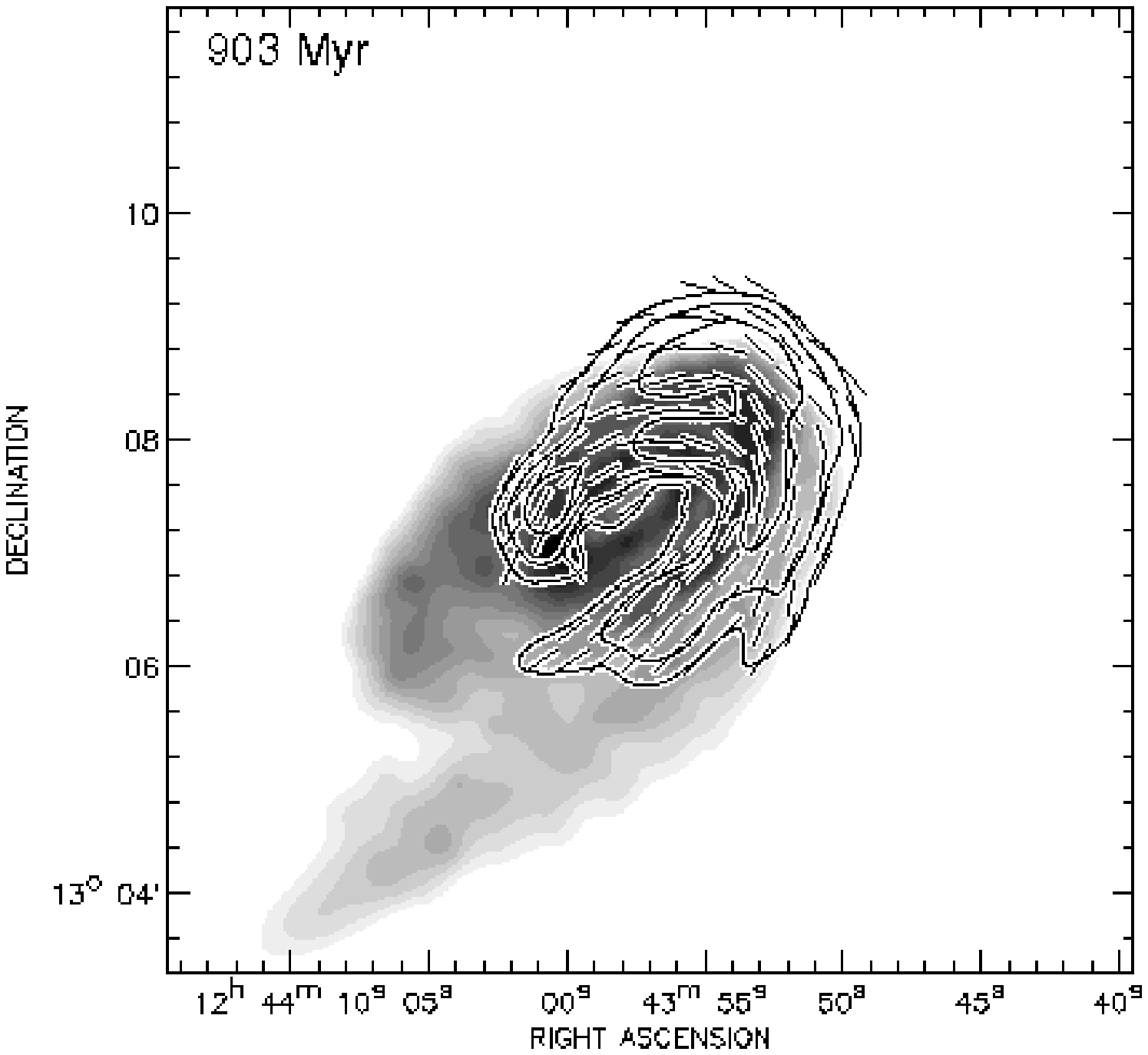}%
\includegraphics[bb=10 40 435 435,width=0.29\textwidth]{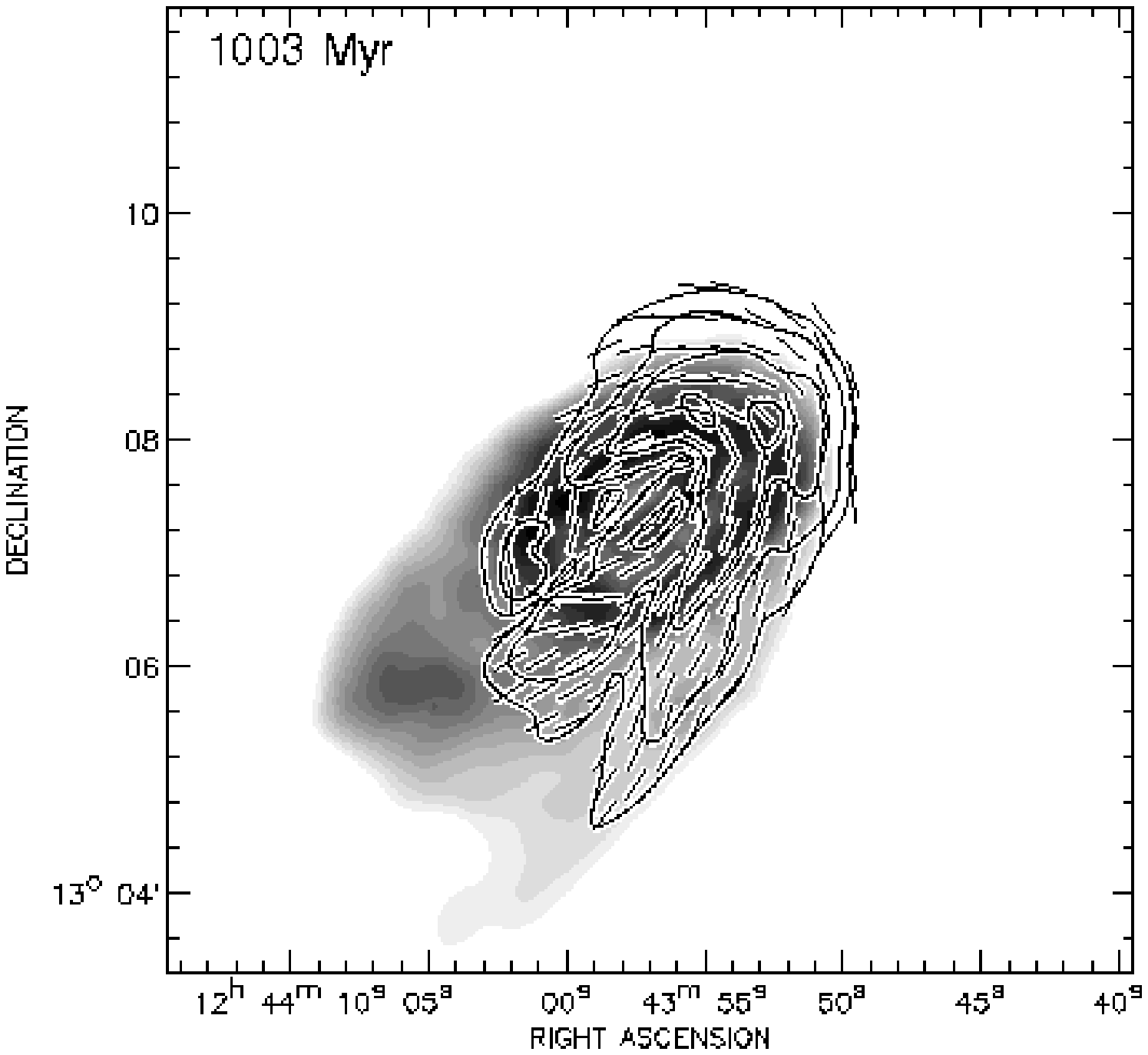}
\caption{
Evolution of the polarized intensity at chosen time steps
for the model with gravitational interaction and 
the constant ram pressure (model GRPS).
Contours: polarized intensity in logarithmic scale.
The polarized intensity
vectors are superimposed onto the gas density distribution in logarithmic scale
(grey plot).
The contours are plotted at levels of (1, 3, 10, 30, 100, 300)
(in arbitrary units) in all plots.
The arrow in the top left panel indicates the direction of the
ram pressure wind; i.e. it is opposite to the galaxy's motion
within the intracluster medium.}
\label{m110}
\end{figure*}

Only when a weak, constant ram pressure is added to the
gravitational interaction between NGC~4654 and NGC~4639
are the observed H{\sc i} distribution and velocity field of
NGC~4654 reproduced (Vollmer~\cite{vol03}).
When a small constant ram pressure is added, the magnetic field
and thus the evolution of the polarized radio-continuum emission
changes considerably (Fig.~\ref{m110}). 
Again, it is possible to compare Fig.~\ref{m110} directly to the snapshots 
of Fig.~14 of Vollmer (\cite{vol03}). 
The polarized intensity maps already show a clear asymmetry at early time steps
because of gas compression due to ram pressure.
Already at $t=300$~Myr the polarized intensity maximum that corresponds to the 
southern maximum of Fig.~\ref{m106} is weaker and displaced to the northwest.
In addition, the polarized intensity maximum that corresponds to the northern maximum
of Fig.~\ref{m106} is brighter due to gas compression and also displaced to the northwest.

Around the timestep of closest approach between the two galaxies ($t=460$~Myr),
tidal forces exceed those due to ram pressure. Therefore, the distribution of
the polarized radio-continuum emission for $450$~Myr$< t < 600$~Myr is
close to that of model GR (Fig.~\ref{m106}). At later time steps, gas compression
due to ram pressure in the northwest and gas re-accretion in the southeast
dominate the gas dynamics. This leads to a pronounced maximum of polarized
emission in the northwest and an elongated ridge of polarized emission
in the south of the galaxy center where the large low column-density
H{\sc i} tail originates (see Phookun \& Mundy~\cite{pm95}).

\subsection{Total magnetic energy evolution}

\begin{figure}[ht]
\centering
\resizebox{\hsize}{!}{\includegraphics[bb=20 30 510 760]{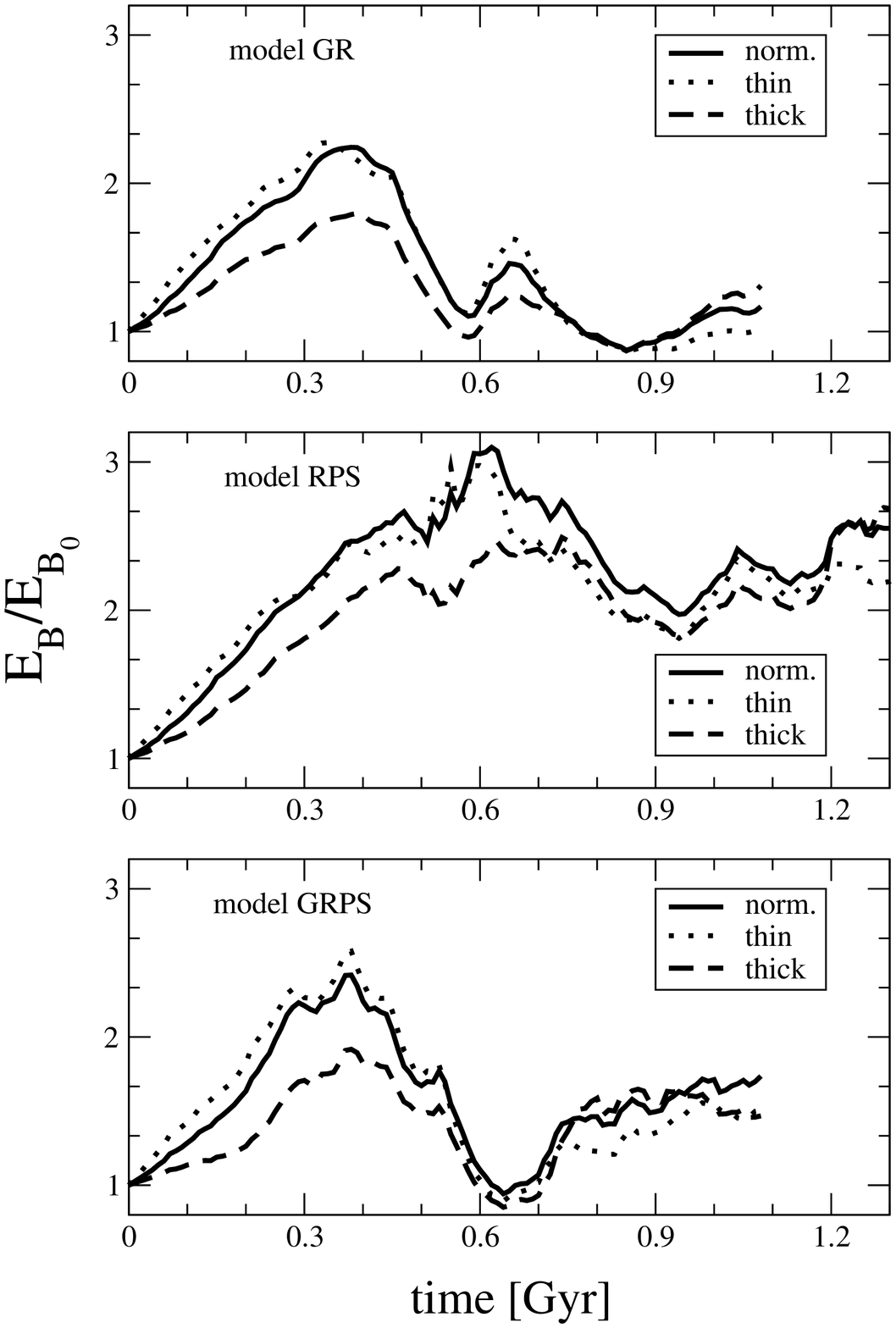}}
\caption{Total magnetic energy evolution in our three models -- ram pressure
alone (model RPS), only gravitational interaction (model GR), and gravitational
interaction with constant ram pressure (model GRPS)}
\label{ene}
\end{figure}

Figure \ref{ene} shows the evolution of the total magnetic energy 
integrated over the whole cube normalized
to its initial value for our three numerical models: GR (tidal interaction
only), RPS (the ram pressure effect alone) and GRPS (tidal interaction and 
constant ram pressure). The magnetic energy is calculated in a cylinder
of 40~kpc diameter and 5~kpc height centered on the galaxy center.

To investigate the impact of the initial conditions of the magnetic field
on the evolution of the total magnetic energy, we used 3 different
initial magnetic field configurations:
\begin{itemize}
\item[(i)]{A ``normal'' disk: the magnetic field is constant up to a radius of 20~kpc,
decreases with a Gaussian profile for
$20$~kpc$< R < 30$~kpc with a half width of 500~pc, and has a cutoff at 30~kpc.
In the vertical direction the initial magnetic field
strength is Gaussian function with a length scale of 500~pc and a cutoff at 1~kpc.}
\item[(ii)]{A ``thin'' disk: the magnetic field is constant up to a radius of 15~kpc,
decreases with a Gaussian profile for
$15$~kpc$< R < 20$~kpc with a half width of 500~pc, and has a cutoff at 20~kpc.
In the vertical direction the initial magnetic field
strength is Gaussian function with a length scale of 300~pc and a cutoff at 600~pc.}
\item[(iii)]{A ``thick'' disk:  the magnetic field is constant up to a radius of 25~kpc,
decreases with a Gaussian profile for
$25$~kpc$< R < 40$~kpc with a half width of 500~pc, and has a cutoff at 40~kpc.
In the vertical direction the initial magnetic field
strength is a Gaussian function with a length scale of 700~pc and a cutoff at 1.5~kpc.}
\end{itemize}

We observe that the evolution of the total magnetic energy of the ``normal'' and
``thin'' disk initial conditions is very close. For $t < 600$~Myr the total magnetic energy of
the ``thick'' disk model is systematically lower than that of the other models;
i.e. the total magnetic energy grows more slowly than for the other two initial conditions.
The reason is that the magnetic field amplification mainly takes place in the inner part of the
galaxy's disk. Therefore, the ratio between the magnetic energy of the amplified field
and that of the initial field is smaller for a larger ``thick'' disk than the same ratio for 
a smaller ``thin'' or ``normal'' disk.
At $t > 700$~Myr, there is no significant difference between the
total energies for the simulations with different initial conditions.
We conclude that the evolution of the total magnetic energy depends only
marginally on the initial configuration of the magnetic field.

The total magnetic energy initially grows in all three models (GR, RPS, GRPS).
This effect can be explained by the magnetic field
amplification due to gradients in the velocity field in early evolutionary
stages. The initial field growth saturates at the time step  of $t \sim 450$~Myr
for models with a gravitational interaction (models GR and GRPS)
and at $t \sim 500$~Myr for model with ram pressure alone (model RPS).
These time steps correspond to the closest encounter with NGC~4639
($t=460$~Myr) and to the timestep of maximum ram pressure ($t=500$~Myr).
For $t>500$~Myr the tidal interaction pulls gas out of the galaxy, taking
the magnetic field with it beyond the limits of our computation box.
Therefore, the total magnetic field energy decreases in models GR and GRPS.  
After the phase of strong interaction ($t \sim 600$~Myr),
a part of the displaced gas falls back onto the galactic disk 
(phase of re-accretion, see Otmianowska-Mazur \& Vollmer \cite{kom03}) 
and its radial flows and shears connected 
to differential rotation of the galactic disk trigger magnetic
field amplification very efficiently.
As a result the magnetic energy of models GR and GRPS increases again at $t=600$~Myr.
At $t>700$~Myr constant ram pressure modifies the gas dynamics creating
shear and a compression region, which leads to a further enhancement of the
total magnetic energy, which then saturates at $t=750$~Myr.

Surprisingly, the enhancement of the total magnetic energy due to
strong ram pressure compression (model RPS) is small ($\sim 10$\%).
The characteristic timescale of the ram pressure stripping event 
(model RPS) is smaller than that of the
gravitational interaction and re-accretion is stronger for model RPS.
The subsequent shear motion makes the large-scale magnetic field increase
at $t>600$~Myr. With the diminishing of re-accretion and thus shear motions 
the magnetic energy decreases again at $t> 800$~Myr and stays approximately
constant for $t>1000$~Myr.

In all models we observe a significant growth in total magnetic energy
at early interaction stages. Only for the model that only includes
ram pressure (model RPS), the total magnetic energy at $\Delta t=t-t_0>300$
(where $t_0$ is the time of strongest interaction) is significantly higher
(a factor of two) than the initial value (see Otmianowska-Mazur \& Vollmer \cite{kom03}).
The stripping of the large scale magnetic field, together with the ISM, might
represent a way to amplify the intracluster magnetic field. 
The observations of ICM show (see eg. Widrow~\cite{wid02}
and references therein) that such fields exist widely in clusters.

We thus obtain a time-dependent global magnetic field amplification in all
our models \emph{without} any explicit magnetic dynamo action
(dynamo coefficient $\alpha=0$).
The amplification was found to work very efficiently in the re-accretion phase
-- when the gas initially pulled out of the galaxy by an external interaction
falls back due to the gravitational attraction by the host galaxy
(see Otmianowska-Mazur \& Vollmer \cite{kom03}).

\section{Models versus observations \label{sec:comparison}}

We first compare our models to the low-resolution Effelsberg data.
This comparison can already distinguish between the different models.
The high-resolution VLA data confirm the choice of the ``best fit'' model.

\subsection{The choice of the model polarized emission threshold}

\begin{figure}[ht]
\centering
\includegraphics[bb=10 40 435 435,width=0.40\textwidth]{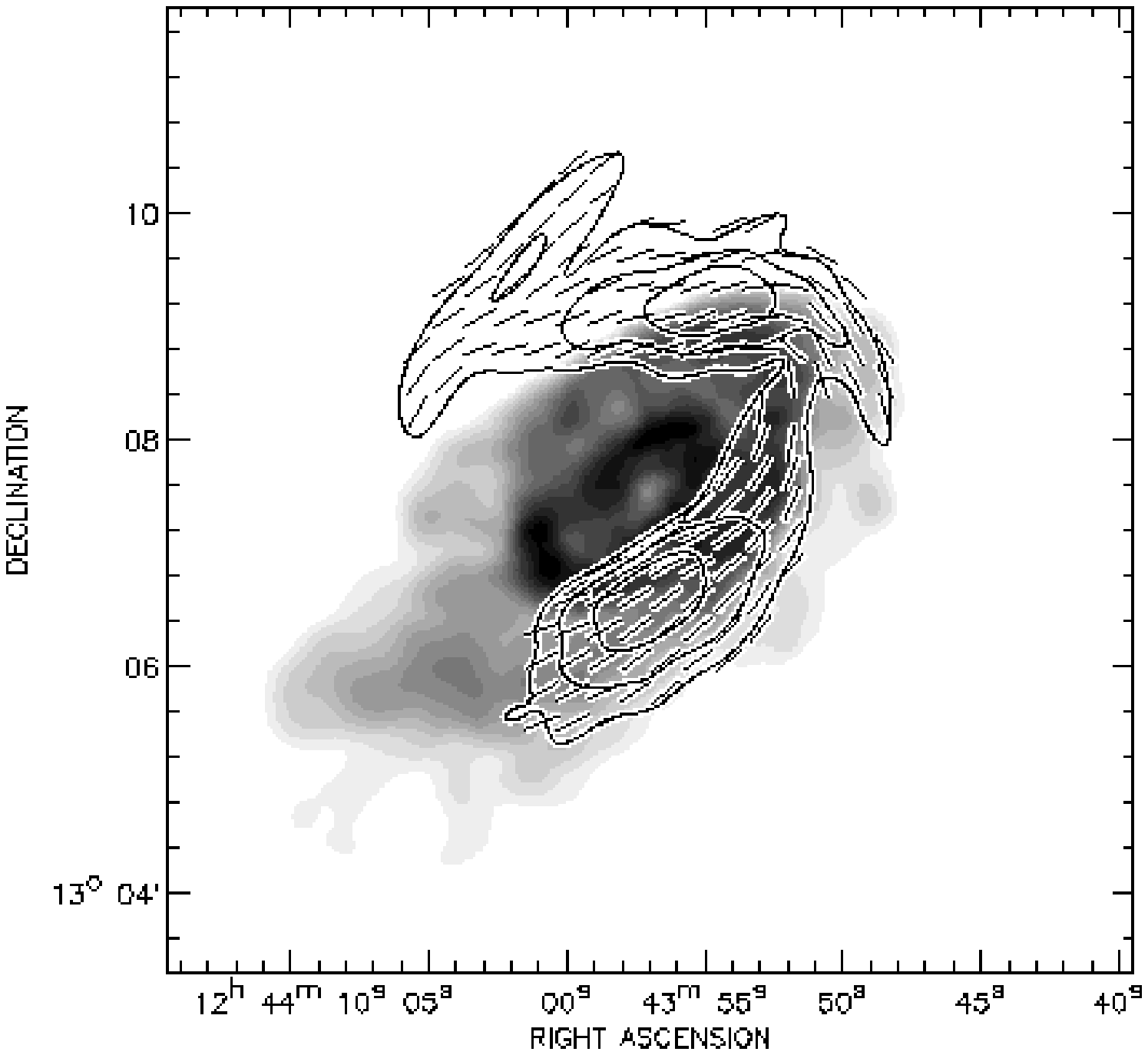}\\
\includegraphics[bb=10 40 435 435,width=0.40\textwidth]{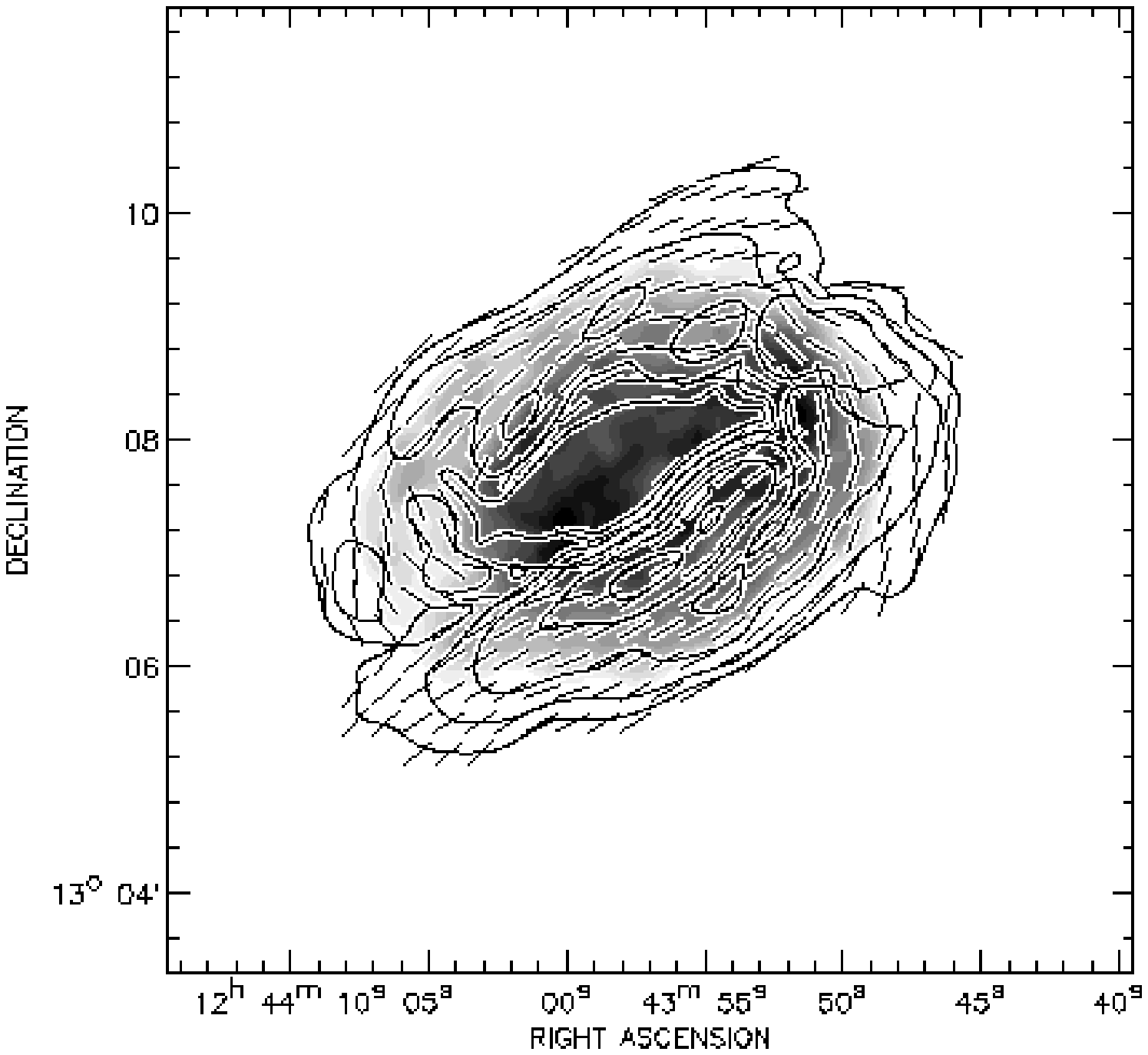}\\
\includegraphics[bb=10 40 435 435,width=0.40\textwidth]{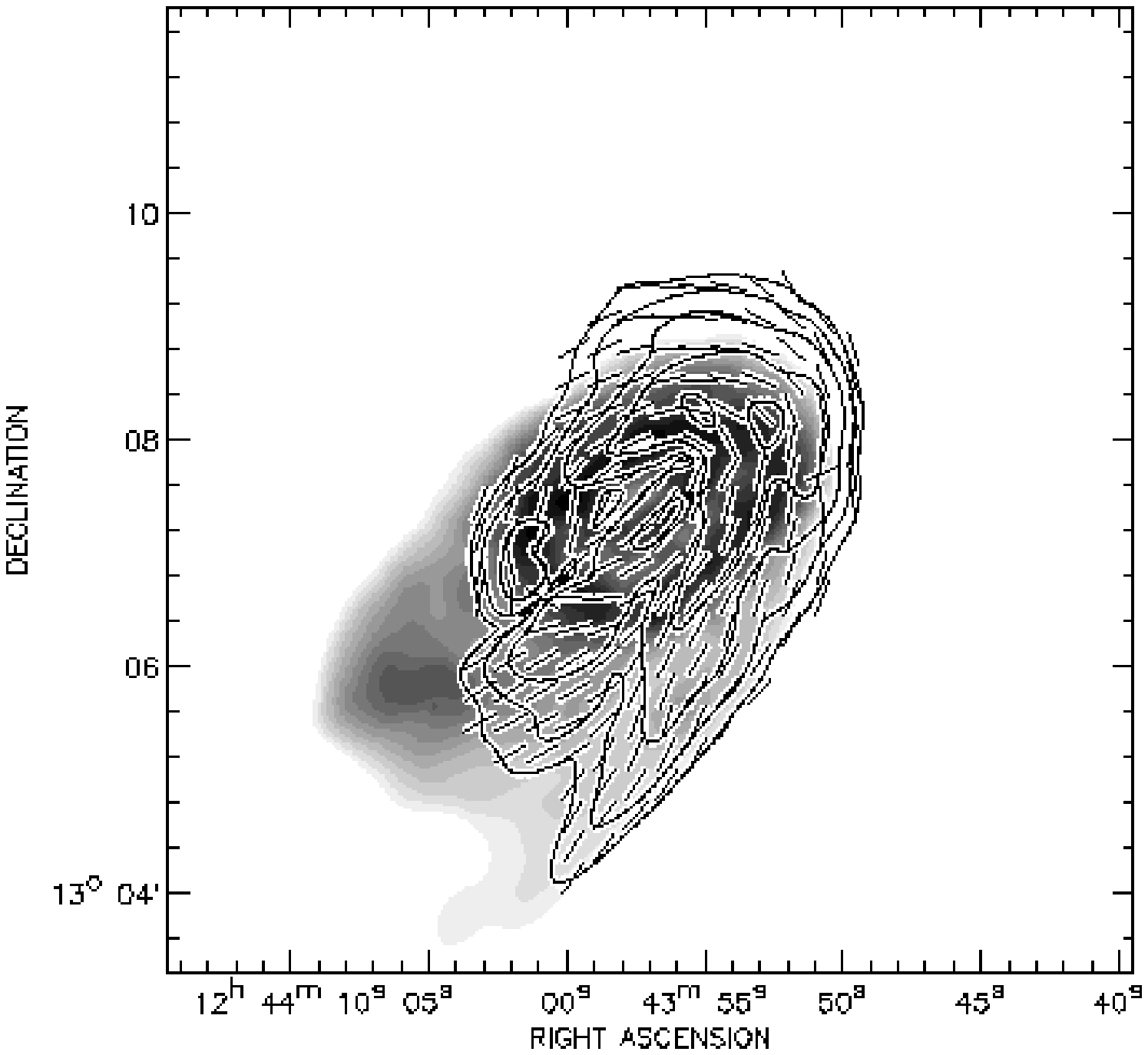}
\caption{Model polarized radio-continuum emission. 
Upper panel: GR, middle panel: RPS, lower panel: GRPS model.
The contour levels
are the same as in Fig.~\ref{m106}, \ref{m086}, and \ref{m110}, 
but the first contour is 0.3 times that of these figures.}
\label{resolution}
\end{figure}

As stated in Sect.~\ref{sec:polmaps} we can only compare the model to the observed polarized 
radio-continuum emission. The model polarized emission is calculated
everywhere in the galaxy according to Eq.~\ref{eq:transfer}. For a proper
comparison we need to know which level of the model polarized emission
corresponds to the observed r.m.s. noise level. Since we do not attempt to estimate the 
absolute synchrotron emissivity $\epsilon_I$, because of our lack of knowledge
about the absolute magnetic field strength and relativistic electron density, we decided to 
use a fixed fraction of the maximum of the model polarized emission distribution
as a model threshold (Fig.~\ref{obs}, \ref{obsvla}).
To give an impression of the polarized radio-continuum distribution at lower
flux densities, we show in Fig.~\ref{resolution} the last snapshots of 
Fig.~\ref{m106}, \ref{m086}, and \ref{m110} with one deeper contour level
at $0.3$ times the lowest contour of the corresponding snapshots. 
For the GR model (upper panel), the two maxima of Fig.~\ref{m106}
are connected by a ridge of faint, polarized emission.
In addition, a faint emission region appears outside the main gas disk
in the north, which is due to very low surface density gas (see Vollmer 2003).
The faint component of the polarized emission distribution of the RPS
model (middle panel) forms a ring structure. Moreover, it extends a little bit
farther out than the 3 times stronger emission.
This is also the case for the faint emission of the GRPS model (lower panel).

As a conclusion, the new polarized emission feature is due to
very low surface-density gas in the case of the GR model.
The lowest contours of the RPS and GRPS models (Fig.~\ref{resolution}) 
(i) do not show any new feature,
(ii) are only a little more extended than the former lowest contours, and
(iii) are a natural extension of the higher contour levels.
Our method is thus not very sensitive to the choice of the model threshold.
The same is true for the convolved data.

\subsection{The Effelsberg data}

\begin{figure*}[ht]
\centering
\includegraphics[width=0.45\textwidth]{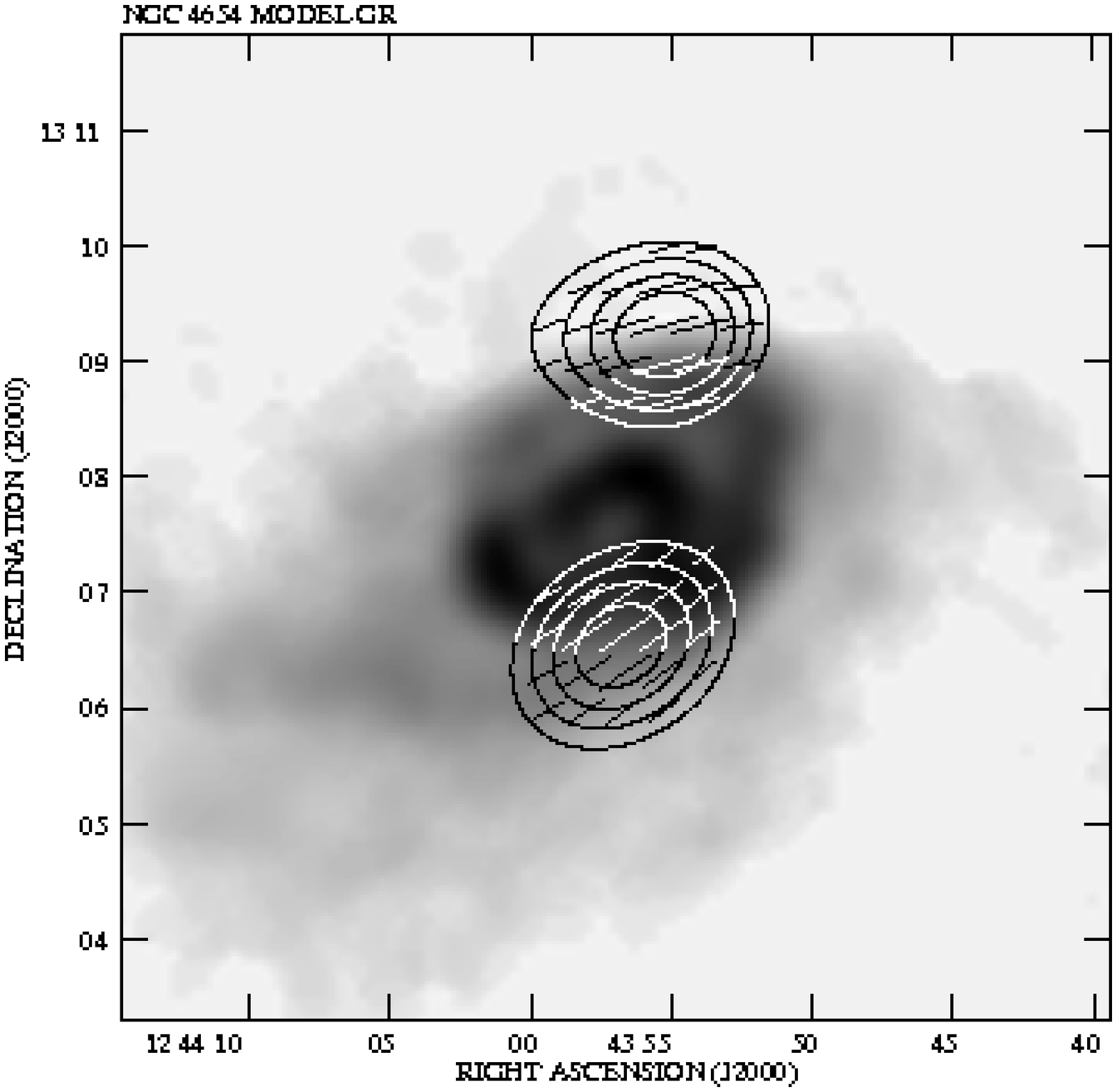}
\includegraphics[width=0.45\textwidth]{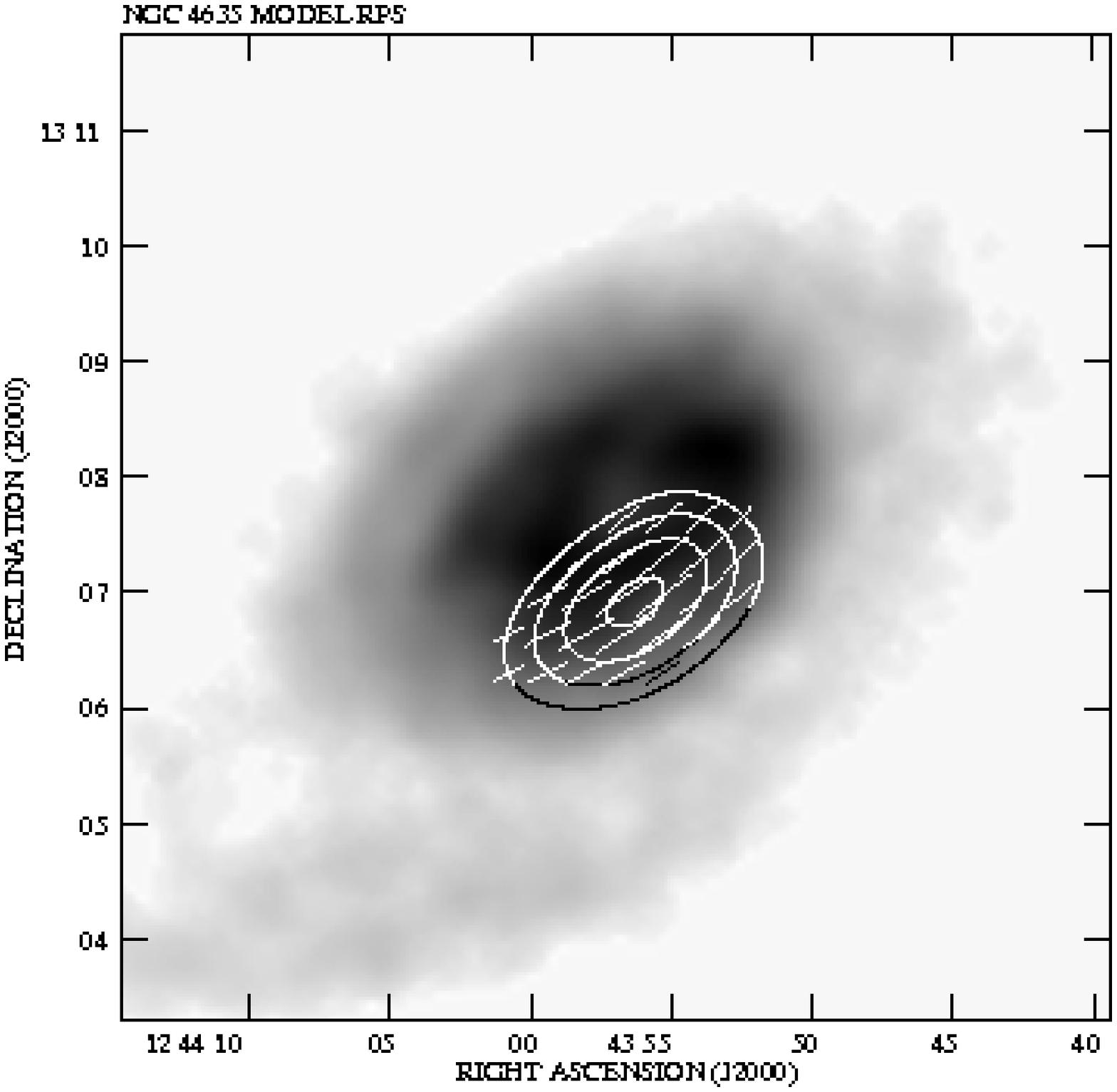}
\includegraphics[width=0.45\textwidth]{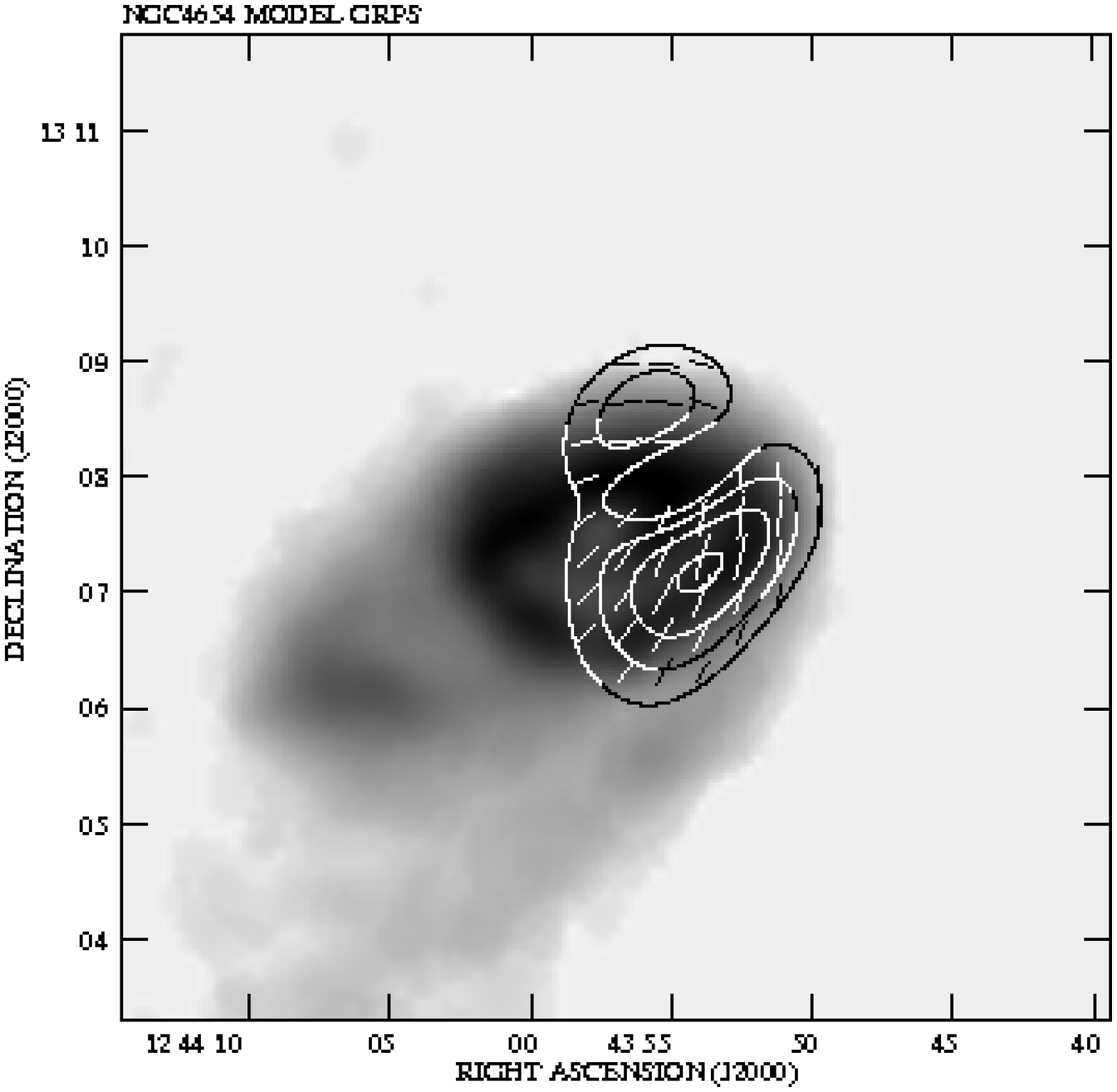}
\includegraphics[width=0.45\textwidth]{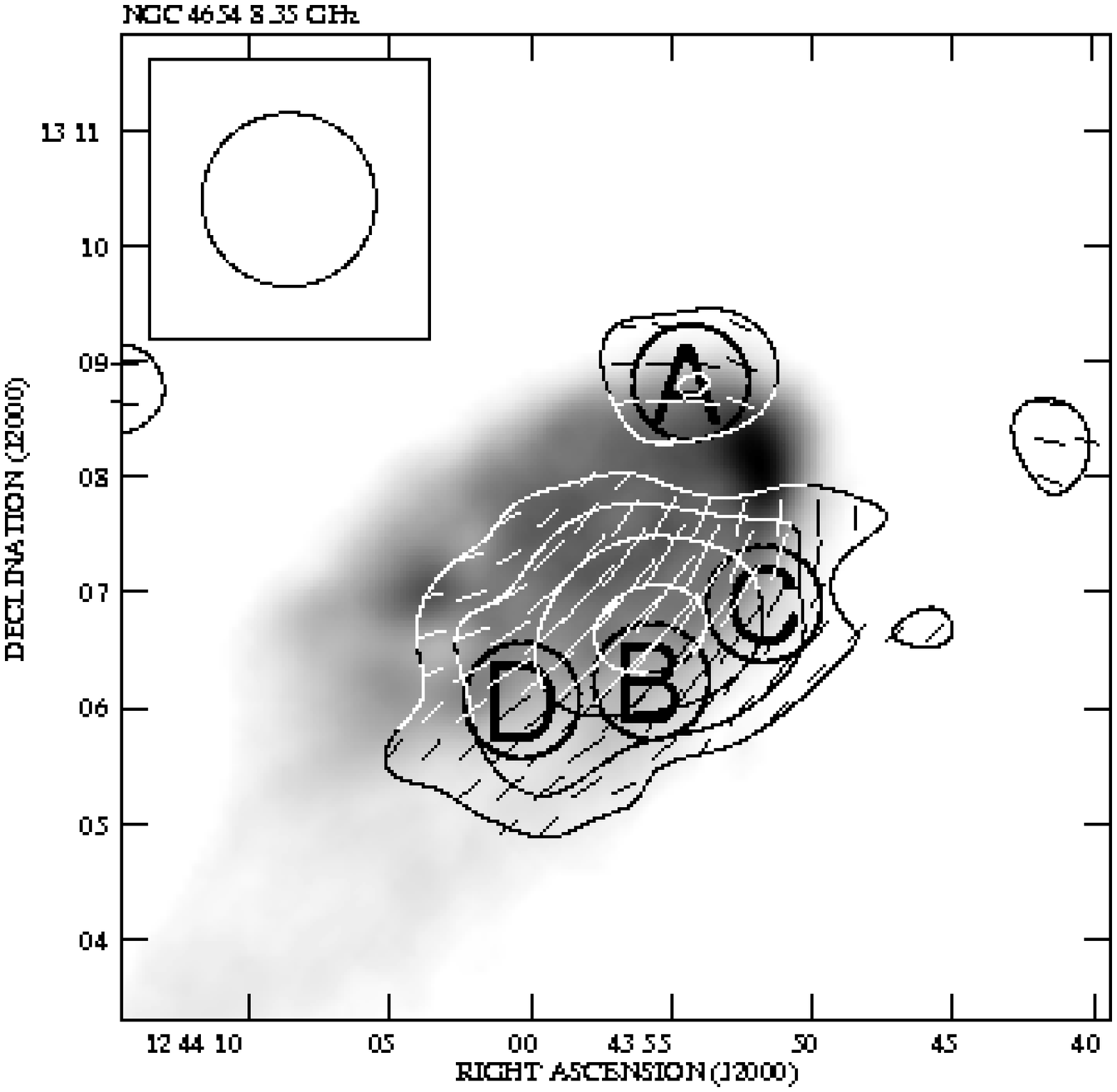}
\caption{Models versus observations for the Effelsberg data. 
Contours: polarized intensity. Grey scale: gas distribution. 
The magnetic field vectors are shown as lines.
{\it Top left}:
model with a gravitational interaction alone (GR). 
Contour levels are (6, 8, 10, 12)$\times$2.5 (in arbitrary units).
{\it Top right}: 
model with ram pressure stripping alone (RPS).
Contour levels are (6, 8, 10, 12)$\times$15 (in arbitrary units).
{\it Bottom left}: model with a tidal interaction and a small 
constant ram pressure (GRPS). 
Contour levels are (6, 8, 10, 12)$\times$10 (in arbitrary units).
{\it Bottom right}:
greyscale: H{\sc i} gas distribution (Phookun \& Mundy \cite{pm95}); 
contours: Effelsberg 8.35~GHz observations.
Contour levels are (3, 5, 8, 12)$\times$40~$\mu$Jy/beam.
}
\label{obs}
\end{figure*}
To compare our simulations to our Effelsberg observations 
(Fig.~\ref{radioobs}), we convolved the final model snapshots to the 
Effelsberg resolution of $80''$ HPBW at 8.35~GHz. 
These convolved snapshots can be seen in Fig.~\ref{obs}
together with our Effelsberg observations.
All three numerical models lead to two maxima of polarized intensity, 
located north and south of the galaxy center. The second maximum of
model RPS cannot be seen, because it is too weak.

There are significant differences in the distribution and strength
of the polarized intensity maxima among our models.
The model only with gravitational interaction (model GR, Fig.~\ref{obs},
top left) shows two maxima of almost equal intensity. 
The model with ram pressure alone (model RPS, Fig.~\ref{obs}, top right)
has only one bright maximum in the south and a very faint maximum
in the north of the galaxy center that is 6.5 times weaker than the
southern maximum.
The simulation including a tidal interaction and a small constant
ram pressure (model GRPS, Fig.~\ref{obs}, bottom left) shows two maxima 
with the northern one 1.3 times weaker than the southern one.
While the locations of the northern maximum are the same for all 
three models, the location of the southern maximum is most eastward
in model GR, westward in model GRPS, and in between in model RPS.

\subsection{The VLA data}

\begin{figure*}[ht]
\centering
\includegraphics[width=0.4\textwidth]{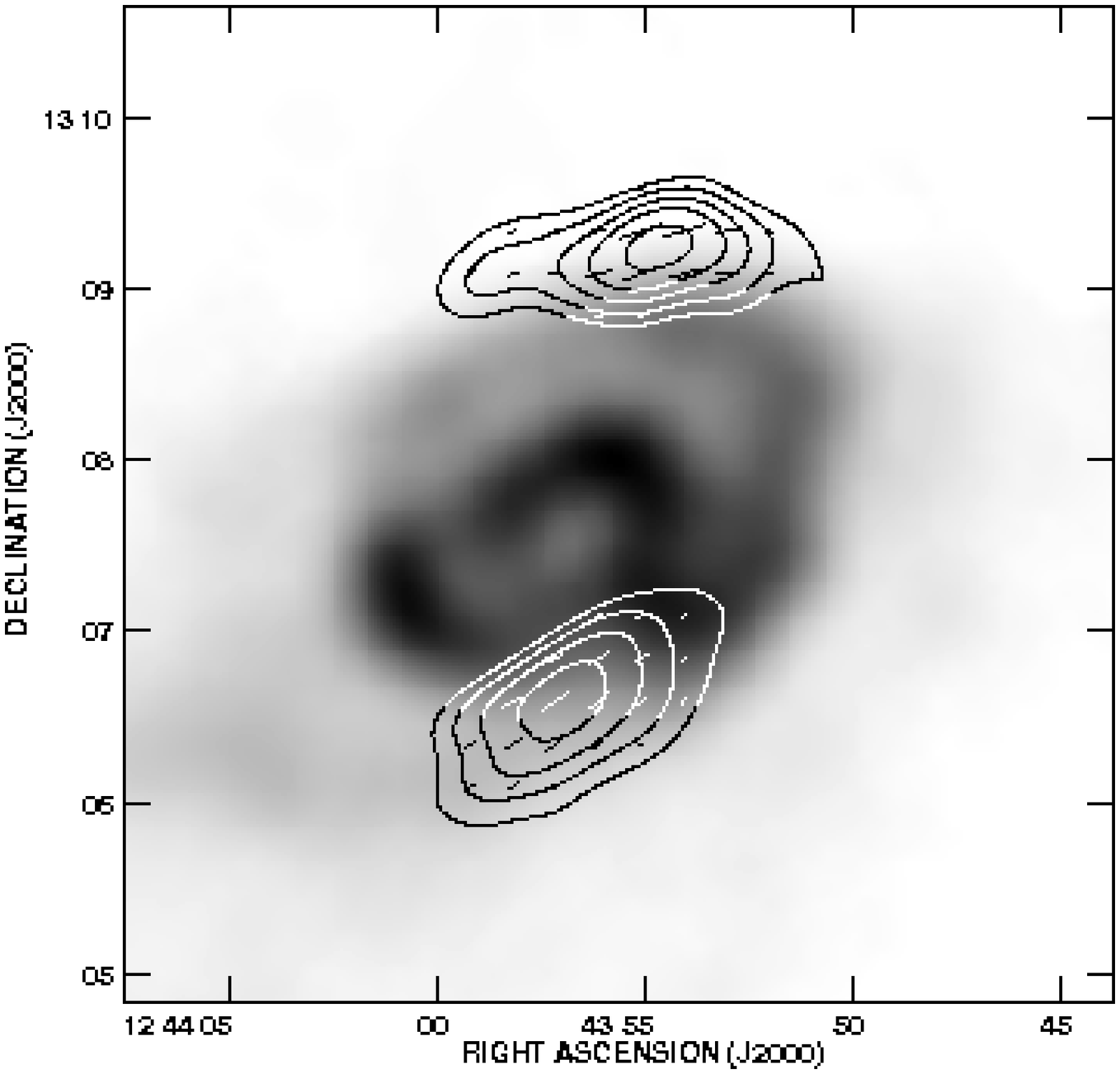}
\includegraphics[width=0.4\textwidth]{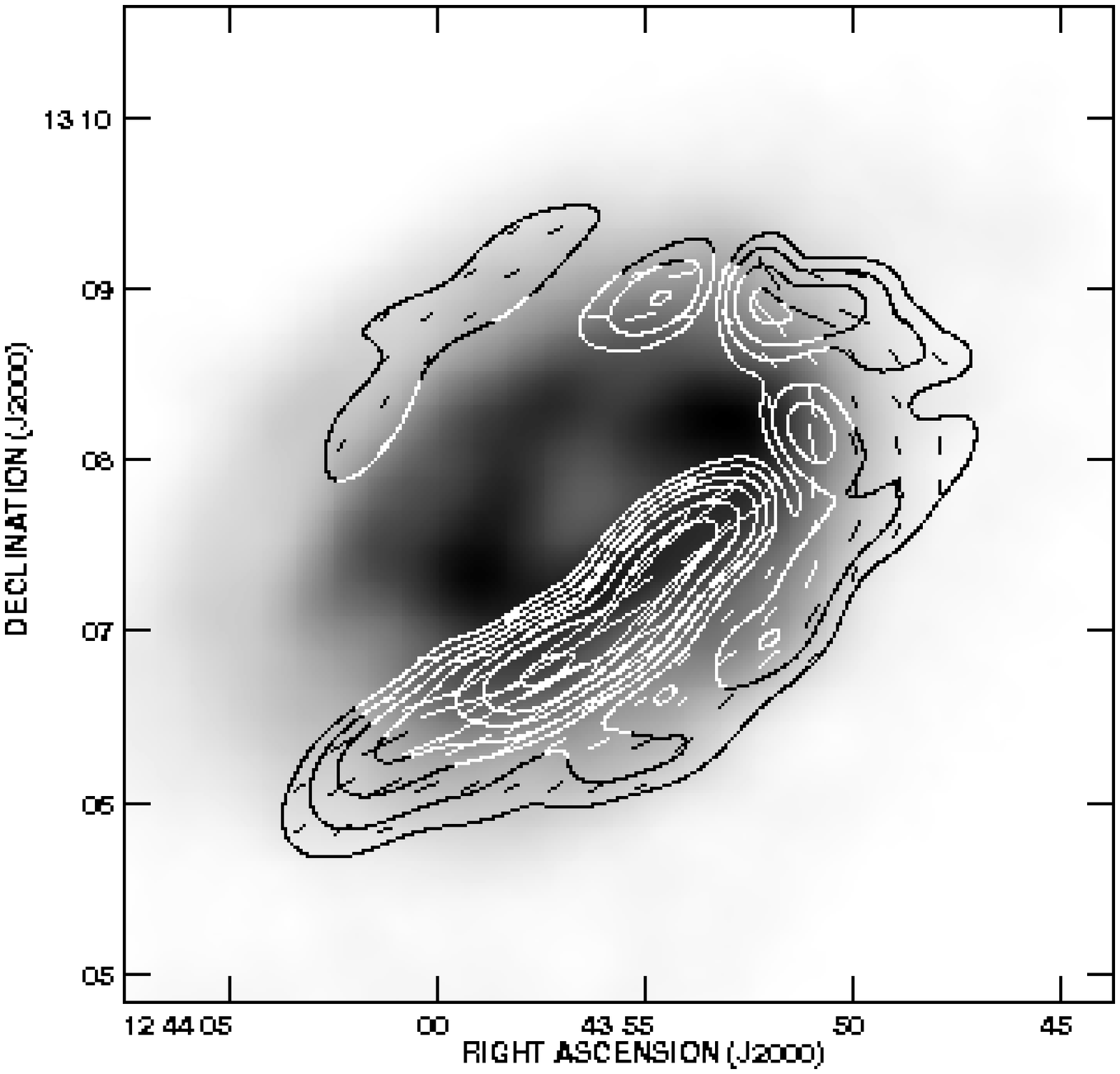}
\includegraphics[width=0.4\textwidth]{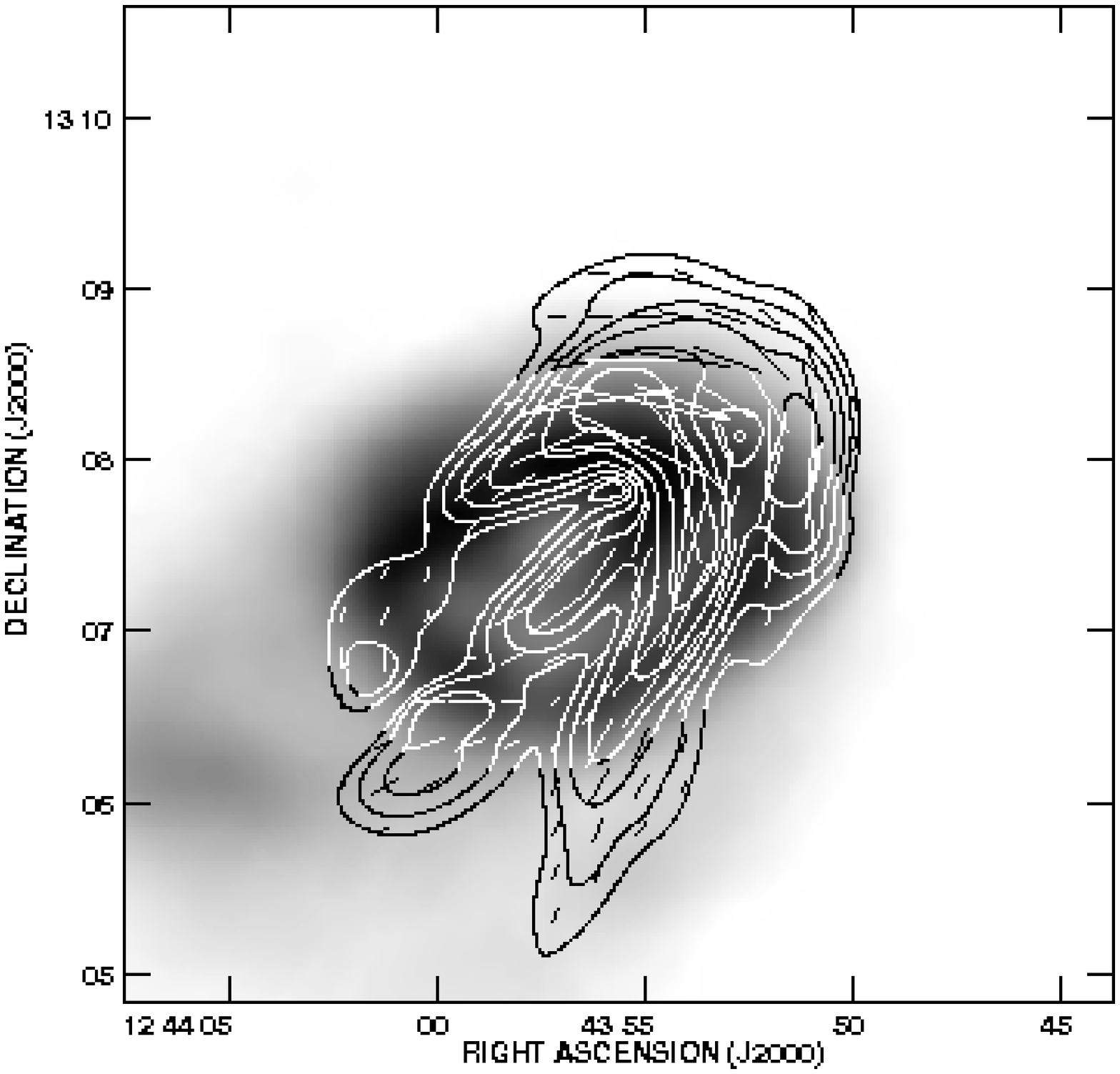}
\includegraphics[width=0.4\textwidth]{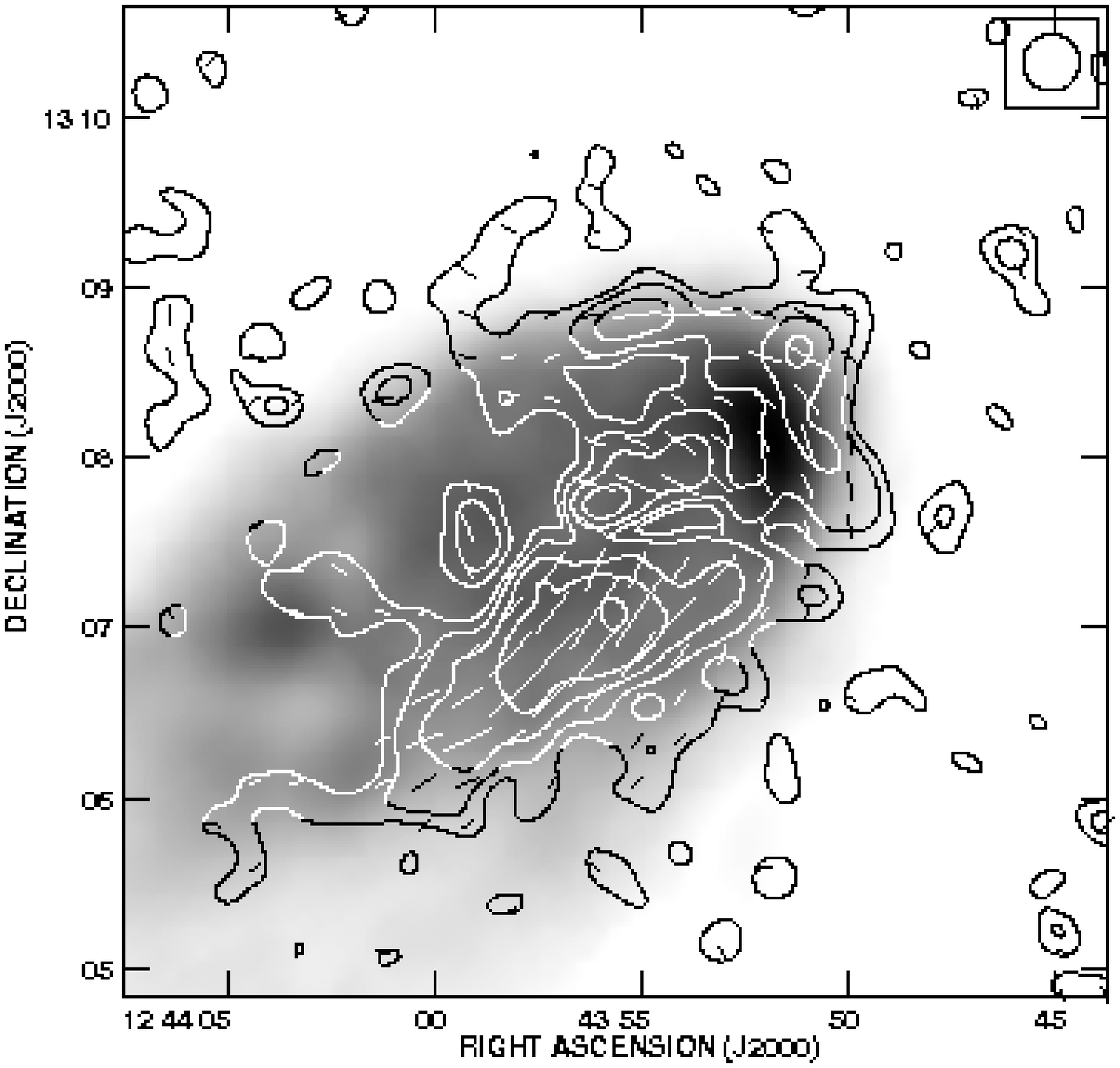}
\caption{Models versus observations for the VLA data. 
Contours: polarized intensity. Grey scale: gas distribution. 
The magnetic field vectors are shown as lines.
{\it Top left}:
model with a gravitational interaction alone (GR). 
Contour levels are (3, 5, 8, 12, 20, 30)$\times$4.3 (in arbitrary units).
{\it Top right}: 
model with ram pressure stripping alone (RPS).
Contour levels are (3, 5, 8, 12, 20, 30)$\times$8 (in arbitrary units).
{\it Bottom left}: model with a tidal interaction and a small 
constant ram pressure (GRPS). 
Contour levels are (3, 5, 8, 12, 20, 30)$\times$8 (in arbitrary units).
{\it Bottom right}:
greyscale: H{\sc i} gas distribution (Phookun \& Mundy \cite{pm95}); 
contours: VLA 4.85~GHz observations.
Contour levels are (3, 5, 8, 12, 20, 30)$\times$8~$\mu$Jy/beam.
}
\label{obsvla}
\end{figure*}

To compare the simulations to our VLA observations 
(Fig.~\ref{radioobsvla}), we convolved the final model snapshots to the 
VLA resolution of $20''$ HPBW at 4.85~GHz. 
These convolved snapshots can be seen in Fig.~\ref{obsvla} together with
our VLA observations.
Since the VLA image has a smaller r.m.s. noise level than the Effelsberg image,
we plot deeper contours in the model polarized radio continuum emission maps.
There is almost no difference between the low and high resolution maps
of the GR model, i.e. there are two distinct maxima in the north and the
south of the galaxy. The low-resolution map of the RPS model only shows the
main ridge of polarized emission, whereas the high resolution map shows
an additional faint ring-like structure.
The difference between the high and low resolution model polarized emission
maps is most significant for the GRPS model where the deep high-resolution map
reveals a complex structure. The main ridge of polarized emission is located in the
west and southwest of the galaxy center. In addition, a distinct maximum can be found
in the north west at the edge of the gas disk.
Polarized emission can be found within the entire southwestern quadrant
of the galactic disk. The southern part of the polarized emission distribution  
splits into two arms in the southeast direction. One arm is located within the 
inner gas disk, the other arm extends in the direction of the extended H{\sc i} tail.
The inner arm is due to the tidal interaction, whereas the outer arm is
mainly caused by ram pressure.

\section{Discussion \label{sec:discussion}}


The GR model produces neither the location of the southern maximum
nor the intensity ratio between the two maxima.
While the location of the southern maximum is reproduced best by model
RPS, the intensity ratio between the two maxima is much too large.
Only the model including a gravitational interaction with NGC~4639
and a constant ram pressure (model GRPS) reproduces 
our Effelsberg observations in a qualitatively satisfactory way. 
Both maxima are placed at about the 
correct positions, and the southern maximum is brighter than the northern one,
although the intensity ratio is still too small (about 1.3 in the model compared to 2.4 in
our observations). 

\begin{table}[ht]
\caption{Mean magnetic pitch-angle values calculated in four chosen regions of
the real galaxy  and our three models' disks (see Fig.~\ref{obs}) --
ram pressure alone (Model RPS), gravitational interaction (Model GR),
and gravitational interaction with constant ram pressure (model GRPS)}
\label{pitch}
\begin{tabular}{cccrrrr}
 &$\alpha_{2000}$&$\delta_{2000}$&NGC&\multicolumn{3}{c}{model}\\
 &$[\;^{\rm h}\;^{\rm m}\; ^{\rm s}\;]$& $[\;^\circ\; ^\prime\; ^{\prime\prime}\;]$&4654&GR&RPS&GRPS\\
\hline
A&12 43 54.3&13 08 48.8& $-$4&    19&     3& $-$9\\
B&12 43 55.7&13 06 13.8&   19&     7&     3&   27\\
C&12 43 51.7&13 06 53.8&$-$18& $-$34& $-$33& $-$7\\
D&12 44 00.3&13 06 03.8&   46&    41&    36&   44
\end{tabular}
\end{table}

As a further step, we compare the magnetic field orientation in our models with 
that of our observations. We calculated the mean values of the pitch angles of
the magnetic vectors in four chosen regions of the galactic disk -- one at the
northern polarized intensity maximum, and three others in the region of the
southern polarized intensity maximum: at the location of the maximum and
on both sides (at a distance of one beam size, see Fig.~\ref{obs}).
These measurements are shown in Table~\ref{pitch}.
The models with a gravitational interaction alone (GR) and with ram pressure
alone (RPS) yield pitch angles too small that are on the southern maximum. 
On the other hand, the pitch angle of the northern maximum of model GR is
too large with respect to the observed value.
The best agreement between the model and observed pitch angles for all
positions is found for model GRPS.


The two prominent maxima of the high-resolution model's polarized emission
map of the GR model are not present in the VLA observations. At first glance,
the overall polarized emission distribution of the RPS model seems
to fit observations quite well. However, the faint northern maximum
is not observed, there is no continuum emission in the inner gas disk,
and the southern maximum is about twice as strong as observed.
Only the GRPS model's polarized emission distribution shows the key
features of the VLA observations: (i) a distinct maximum can be found
in the northwest at the edge of the gas disk.
This maximum is due to gas compression by ram pressure; (ii)
polarized emission can be found within the entire southwestern quadrant
of the galactic disk; (iii) there is a region of strong polarized emission
in the southwest of the galaxy. However, we do not observe the bifurcation
of the polarized emission ridge to the southeast. Especially the arm
pointing toward the extended H{\sc i} tail is missing in the observations.
This might be due to a difference between the model and real distribution 
of relativistic electrons. Despite this disagreement, we feel confident that
the GRPS model reproduces our VLA observations best and thus confirms our
findings based only on the Effelsberg data.


We conclude that GRPS model not only successfully reproduces the
H{\sc i} distribution and velocity field, but also successfully reproduces the
observed polarized radio-continuum emission distribution.
Observations of polarized emission can serve as an additional
diagnostic tool for galactic interactions
(see Otmianowska-Mazur \& Vollmer \cite{kom03}). It can add the information
on gas flows perpendicular to the line of sight, adding a new dimension
to radial velocity studies (see also Urbanik~\cite{urb05}).

In the case of NGC~4654 we identified the northwestern polarized emission
maximum as due to ram pressure. As in NGC~4522 (Vollmer et al. \cite{vol06}), we can
detect the compression of the interstellar medium directly in polarized
radio-continuum emission. This independently confirms the claim of
Phookun \& Mundy (\cite{pm95}) that the linear H{\sc i} rotation curve in
the northwestern part of the galaxy is due to compression and the
same conclusion of Vollmer (\cite{vol03}) which was based on the comparison
between deep H{\sc i} data and a dynamical model. In the case of
NGC~4522 (Vollmer et al. \cite{vol06}), it was not even possible to
detect the compression region in the H{\sc i} velocity field.
A part of the southern polarized radio continuum emission ridge is due to
the tidal interaction, which is also responsible for the one-armed spiral.
Another part is due to shear motions induced by the action of ram pressure.
It is very difficult to determine regions of shear motions only on the basis
of a radial velocity field. Since the polarized radio continuum emission is
sensitive to the 3D velocity field, one can gain
important information about transversal motions. By definition, this
information is not contained in a radial velocity field. 
Thus, polarized radio continuum emission gives us important information
about shear and compression regions that may not be accessible via
velocity fields.

Our present study corroborates the results of Vollmer (\cite{vol03}), that
NGC~4654 had a recent close and rapid tidal interaction with NGC~4639.
Cepheid measurements (Gibson et al. \cite{gib00}, Freedman et al. \cite{fre01})
place NGC~4639 several Mpc behind the Virgo cluster center (M87, see Sect.~1).
However, H-band Tully-Fisher measurements (Gavazzi et al. \cite{gav99})
yield a distance of NGC~4654 of 1-2~Mpc in front of the Virgo cluster
center. In light of our new results, we suggest that
the Tully-Fisher distance is underestimated and that the line-of-sight 
distance of NGC~4654 is close to that of NGC~4639. 

Another result of our study is that a small amount of ram pressure
is needed to reproduce the H{\sc i} gas distribution, velocity field,
and the distribution of the polarized radio-continuum emission.
An intracluster medium (ICM) density of $\sim 10^{-4}$~cm$^{-3}$ is
needed for this ram pressure (see Sect.~1). If the ICM density distribution is
spherical, the density at a distance of 5~Mpc is more than a magnitude
smaller. A density of $10^{-4}$~cm$^{-3}$ is already reached at a distance of 
1~Mpc from the cluster center (Schindler et al. \cite{sch99}).
We therefore suggest that the ICM
distribution of the Virgo cluster is not spherical but elongated in
the line of sight with an axis ratio of about 5:1 as observed for the
galaxy distribution (Gavazzi et al. \cite{gav99}, Yasuda et al. \cite{yas97}).

\section{Conclusions \label{sec:conclusions}}

The Virgo spiral galaxy NGC~4654 was observed with the Effelsberg 100m
telescope at 8.35~GHz and the VLA at 4.85~GHz in polarization. 
We find an asymmetric distribution
of the polarized intensity with a strong peak in the south of the galaxy. 
This shows that asymmetric polarized radio continuum distributions 
are typical of cluster spiral galaxies (Vollmer et al. \cite{vol04}, Urbanik \cite{urb05}).

The polarized radio continuum emission was used as an additional
diagnostic tool for the interaction between NGC~4654 and its environment.
To do so, we solved the induction equation for the dynamical model evolution 
of the interstellar medium of a cluster spiral galaxy and calculated the
large-scale magnetic field in this way. With an assumed distribution of
relativistic electrons, the distribution of polarized radio-continuum emission
can be calculated and directly compared to our observations.
The input velocity fields are those from the simulations of
Vollmer (\cite{vol03}),
who simulated three different interaction scenarios: (i) a gravitational
interaction between NGC~4654 and NGC~4639, (ii) ram pressure stripping
by the intracluster medium, and (iii) a gravitation interaction, along 
with a small constant ram pressure. 

The direct comparison between our MHD simulations and the polarized 
radio-continuum observations shows that only the model with gravitation 
interaction and low, constant ram pressure can reproduce 
observations in a satisfactory way. We found that the low-resolution Effelsberg
data are sufficient for distinguishing between the models.
Thus, our study corroborates the results of Vollmer et al.
(\cite{vol04}). This mixed interaction succeeds in reproducing
the H{\sc i} gas distribution and velocity field (Vollmer \cite{vol03})
and the polarized radio-continuum distribution.

Since Cepheid measurements place NGC~4639 $\sim 5$~Mpc behind the center
of the Virgo cluster, we suggest that the distribution of the Virgo 
intracluster medium is highly elongated thereby providing the
necessary ram pressure.
Thus, polarized radio-continuum emission, in combination with detailed MHD
modeling, provides an important additional diagnostic tool for studying
interactions between a galaxy and its cluster environment.
This tool is complementary to deep H{\sc i} observations.

\acknowledgements
This work was supported by the Polish-French (ASTRO-LEA-PF) cooperation program,
and by the Polish Ministry of Sciences grant PB 378/P03/28/2005.

\end{document}